\documentclass[10pt]{article}

\pdfoutput=1 

\usepackage{jheppub} 

\usepackage[T1]{fontenc} 
\usepackage{enumitem}

\usepackage{amssymb,amsthm,cancel,hyperref,graphicx,xcolor}
\usepackage{mathrsfs,wasysym}
\usepackage{booktabs}
\usepackage{tikz}
\usetikzlibrary{calc}
\usepackage{placeins}

\newcommand{\gsim}{\gtrsim}
\newcommand{\lsim}{\lesssim}

\newcommand{\MSbar}{\overline{\text{MS}}}

\newcommand{\as}{\alpha_s}

\newcommand{\Ord}{\mathcal{O}}

\newcommand{\Li}{\mathrm{Li}}
\newcommand{\gammae}{\gamma_{\scriptscriptstyle E}}

\let\originalleft\left
\let\originalright\right
\renewcommand{\left}{\mathopen{}\mathclose\bgroup\originalleft}
\renewcommand{\right}{\aftergroup\egroup\originalright}

\newcommand{\zc}{z_{\rm cut}}

\newcommand{\tausd}{\tau_\textsc{\tiny SD}}

\newcommand{\herwig}{\textsf{Herwig}}
\newcommand{\pythia}{\textsf{Pythia}}
\newcommand{\sherpa}{\textsf{Sherpa}}
\newcommand{\fastjet}{\textsf{FastJet}}

\newcommand{\eventtwo}{\textsf{EVENT2 }}

\newcommand{\new}{}

\def\beq{\begin{equation}}  
\def\eeq{\end{equation}}
\def\({\left(}
\def\){\right)}
\def\[{\left[}
\def\]{\right]}

\title{\boldmath Soft-Drop Thrust}

\author[a]{Jeremy~Baron,}
\author[b]{Simone~Marzani,}
\author[c]{Vincent~Theeuwes.}

\affiliation[a]{University at Buffalo, The State University of New York, Buffalo, NY 14260-1500, USA}
\affiliation[b]{Dipartimento di Fisica, Universit\`a di Genova and INFN, Sezione di Genova,\\ Via Dodecaneso 33, 16146, Genoa, Italy}
\affiliation[c]{Institute for Theoretical Physics, Georg-August-Univesity G\"ottingen,\\ Friedrich-Hund-Platz 1, 37077 G\"ottingen, Germany}
\emailAdd{jfbaron@buffalo.edu}
\emailAdd{simone.marzani@ge.infn.it}
\emailAdd{vincent.theeuwes@uni-goettingen.de}

\abstract{
Soft drop, a technique originally developed in the context of jet physics in proton-proton collisions in order to reduce the contamination from non-perturbative effects, is applied to event shapes in electron-positron annihilation.
In particular, we study the thrust distribution at the $Z$ pole and show that the region where non-perturbative corrections due to the hadronisation process are small is considerably extended if soft drop is applied.
Therefore, we argue that the use of soft drop to reduce hadronisation effects is potentially of great benefit in the context of strong coupling determination using event shapes, which would be otherwise characterised by a strong correlation between $\alpha_s$ and non-perturbative parameters. 
However, reduced sensitivity to hadronisation corrections is only one of the aspects that need to be considered. In this context, we show that perturbative calculability, especially away from the soft and collinear region of the event-shape spectrum, has a nontrivial interplay with the soft-drop observable of choice. To this purpose, besides thrust, we investigate the behaviour of the hemisphere mass as well as the jet mass.
We find that the latter shows the most promising behaviour in the intermediate region of the spectrum, especially if small jet radii are considered.
}

\usepackage{graphicx}
\begin{document}
\maketitle

\section{Introduction}

In the absence of a striking signature of new physics, the success of the physics programme of the CERN Large Hadron Collider (LHC) heavily relies on our ability to perform theoretical calculations with ever decreasing uncertainties and compare them to precise experimental data, in order to achieve a deeper knowledge of the Standard Model of particle physics and, eventually, to find evidence of deviation from it. 
In this enterprise, high-precision calculations in perturbative QCD play a central role and indeed a lot of theoretical effort has been put into performing calculations both at fixed-order and resummed levels. In order to achieve the sought-after accuracy, this effort must be accompanied by reliable determinations of the parameters that enter the Standard Model Lagrangian, such as masses and couplings. 
In particular, the relative size of perturbative QCD corrections is determined by the strong coupling constant, $\as$, and precision QCD requires a reliable determination of this parameter. 
The current value of the strong coupling determined by the Particle Data Group is  $\as(m_Z)=0.1181 \pm 0.0011$~\cite{Patrignani:2016xqp}.\footnote{The size of the uncertainty of the world average, which is actually an average of averages, is doubled with respect to its previous determination $\as(m_Z)=0.1185 \pm 0.0006$~\cite{Agashe:2014kda}, mostly because of a more conservative treatment of the uncertainty that affects lattice QCD calculations.}
One type of observables that is traditionally employed in $\alpha_s$ extraction using perturbative QCD are event shapes in electron-positron ($e^+e^-$) collisions. 

Event-shapes quantitatively describe final-state QCD radiation and therefore provide a rather clean way of exposing the strong coupling. On the one hand, differential distributions for these types of observables are known to next-to-next-to-leading order (NNLO) in $\alpha_s$~\cite{GehrmannDeRidder:2007jk,Gehrmann-DeRidder:2007nzq,GehrmannDeRidder:2007hr,Weinzierl:2009ms,Weinzierl:2010cw,DelDuca:2016ily} and are therefore used in precision determinations of the strong coupling. On the other hand, unless one imposes rather stringent cuts on the value of the event-shape, these distributions acquire sensitivity to the emission of soft and collinear partons, which results in potentially large logarithmic corrections. Therefore, state-of-the-art determinations of event shapes combine together fixed-order calculations with the all-order resummation of these large corrections. Next-to-leading logarithmic (NLL) resummation for specific event shapes have been known for a while, e.g.\ \cite{Catani:1992ua}, and a framework to resum a rather general class of event shapes also exists~\cite{Banfi:2010xy,Banfi:2001bz,Banfi:2004yd,Banfi:2003je,Banfi:2004nk}. In recent years, this framework has also been extended to NNLL~\cite{Banfi:2014sua,Banfi:2016zlc}. Furthermore, dedicated resummed calculations have been performed both in context of direct QCD resummation and using the methods of Soft-Collinear Effective Theory (SCET), e.g.\ \cite{Becher:2008cf,Monni:2011gb}. 
These high-precision calculations have been extensively used in the context of $\alpha_s$ fits to experimental data, e.g.\ \cite{Dissertori:2007xa,Davison:2008vx,Bethke:2008hf,Dissertori:2009ik,OPAL:2011aa}.

However, some of the high-precision determinations of $\alpha_s$ significantly differ from the world average. A striking example is provided by fits performed using the event shapes thrust~\cite{Farhi:1977sg} and $C$-parameter~\cite{Fox:1978vu}. The calculation used in these determinations is of an astonishing theoretical precision: it includes resummation to N$^3$LL matched to NNLO. 
The obtained value is $\as(m_Z)=0.1135\pm 0.0011$, which is a few standard deviations below the world average~\cite{Abbate:2010xh,Hoang:2015hka}. 
This rather surprising result clearly demands further investigation.
Given the fact that from a perturbative viewpoint, these calculations represent the state of the art for both fixed-order and resummed calculations, it is natural to put non-perturbative corrections under scrutiny.
Thrust is an infra-red and collinear (IRC) safe observable and thus one expects non-perturbative corrections to be suppressed by inverse powers of the hard scale $Q$. Nevertheless, hadronisation corrections at the energies of interest, i.e the LEP centre-of-mass energy, turn out to be sizeable. Their primary effect is a shift in the position of the peak of the thrust distribution, together with a distortion of the spectrum in the peak region. In the approach of Refs.~\cite{Abbate:2010xh,Hoang:2015hka}, hadronisation corrections are taken into account by fitting a universal one-parameter non-perturbative soft function defined in field theory. However, it turns out that this non-perturbative parameter is strongly correlated with $\as$ and thus the simultaneous fit of the two has some degree of degeneracy.

A possible way out would be to consider measurements of multiple observables, possibly at different centre-of-mass energies, in order to break the degeneracy between perturbative and non-perturbative physics. However, this requires having high-precision predictions for multiple observables and while NNLO calculations can be performed at the fully differential level, resummed distributions often require dedicated calculations. Furthermore, one should probably go beyond the rather simple one-parameter modelling of the hadronisation process.
In this paper, we put forward a different approach and we begin to explore its feasibility, at least from a theoretical point of view. Namely, we suggest that rather than looking for a way to disentangle perturbative and non-perturbative physics, we should focus on observables that, while maintaining several features of the commonly used event shapes, have at the same time reduced sensitivity to non-perturbative corrections. One way of constructing such observables is through the application of so-called grooming algorithms, which have been developed in the context of jet physics at the LHC.

The field of jet substructure~\cite{Abdesselam:2010pt,Altheimer:2012mn,Altheimer:2013yza,Adams:2015hiv,Larkoski:2017jix} aims to develop efficient ways to distinguish signal jets originating from the decay of highly-boosted massive particles into hadrons, from the overwhelming background of QCD jets. In particular, many jet substructure algorithms contain a grooming step, namely a procedure to remove soft and large-angle radiation from the jet, as this is likely to come from contamination with the busy environment that one encounters in proton-proton collision. Grooming algorithms decrease, by construction, the effective radius of a jet and, therefore, its area~\cite{Cacciari:2008gn}, thus reducing  the sensitivity of jet observables from the underlying event and pile-up. The effect that these algorithms have on hadronisation corrections depends instead on the algorithm of choice~\cite{Dasgupta:2013ihk}. However, Monte Carlo studies show that the widely used (modified) Mass-Drop Tagger (mMDT)~\cite{Butterworth:2008iy,Dasgupta:2013ihk}, trimming~\cite{Krohn:2009th}, pruning~\cite{Ellis:2009su,Ellis:2009me}, and soft drop~\cite{Larkoski:2014wba}, all exhibit reduced sensitivity to non-perturbative hadronisation corrections.
In this list, the mMDT/soft-drop algorithms are the best understood from a theoretical viewpoint.
Indeed, significant progress has been made to perform all-order calculations for soft-drop observables~\cite{Marzani:2017mva,Marzani:2017kqd}. In the context of SCET, computations have been performed up to NNLL accuracy~\cite{Frye:2016aiz,Frye:2016okc} for the soft-drop mass (see also~\cite{Kang:2018jwa}) and, more recently, for multi-prong jet shapes~\cite{Larkoski:2017cqq}. 
Unfolded measurements also exist~\cite{CMS:2017tdn,Aaboud:2017qwh}, which show very good agreement with perturbative predictions. 

The soft-drop algorithm is a powerful tool that reduces the sensitivity of jet observables to non-perturbative contributions, such as hadronisation and the underlying event, thus extending the domain of applicability of high-precision perturbative calculations in QCD. It is therefore natural to explore its application to QCD final states in $e^+e^-$ collision, where the only non-perturbative contribution arises from the hadronisation process, with the aim of reducing their impact. This is what we are set to do in this study. 
In the first part of this paper, we study the impact of non-perturbative corrections on the thrust distribution at LEP energies. In particular, in Section~\ref{sec:jet-substr} we define soft-drop event shapes, while we perform a detailed Monte Carlo study in Section~\ref{sec:MC}. 
In the second part of the paper, in view of using soft-drop event shapes for future extractions of the strong coupling, we study the interplay of the soft-drop algorithm with perturbative predictions, discussing both resummation and fixed-order. We consider the thrust distribution in Section~\ref{Sec:Analytics}, while we discuss the jet mass in Section~\ref{sec:Jet-mass}. Finally we conclude in Section~\ref{sec:conclusion}. Explicit results and technical details are collected in the Appendices.

\section{Thrust with soft drop}\label{sec:jet-substr}
\begin{figure}
	\centering
	\includegraphics[width=0.49\textwidth]{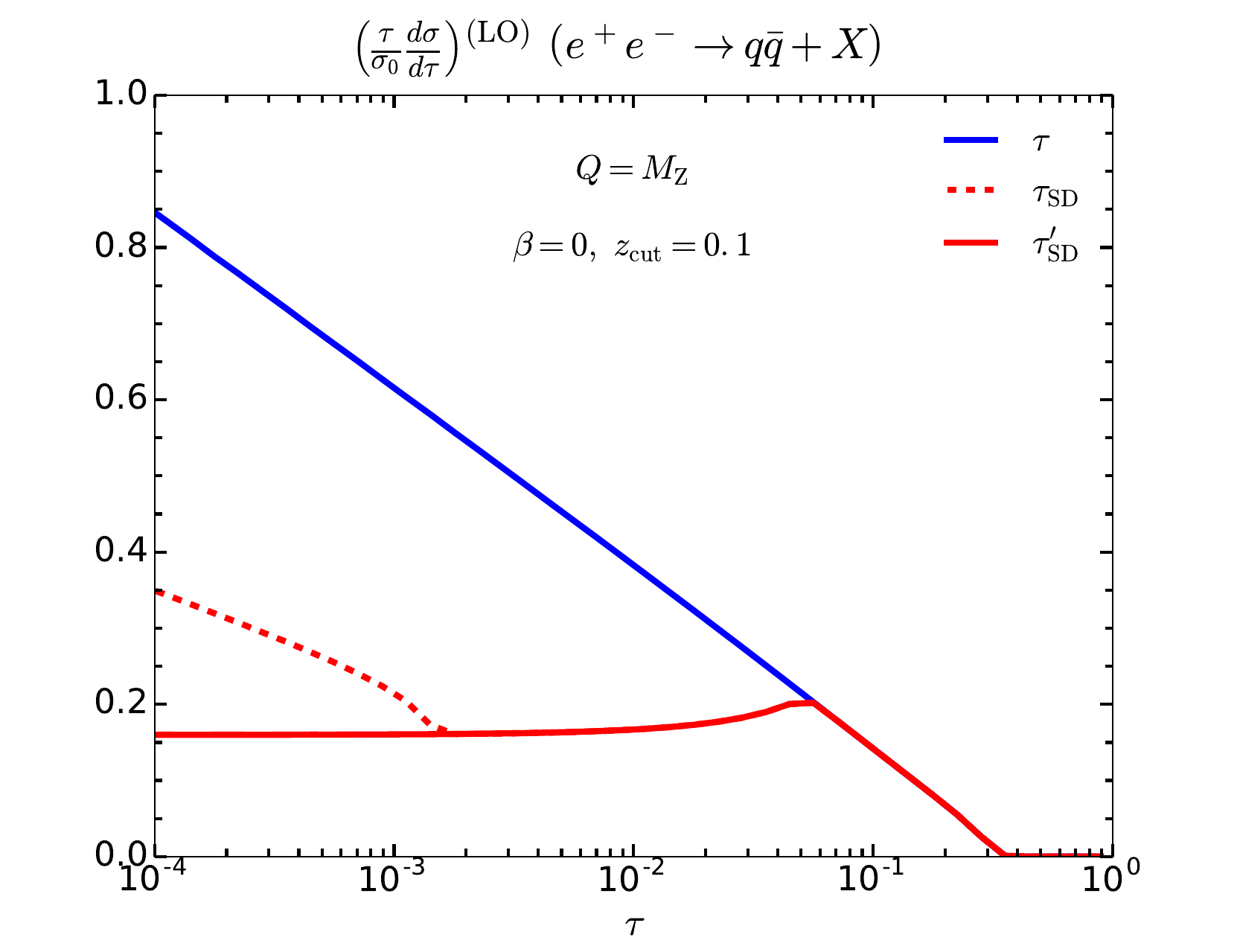}
         \includegraphics[width=0.49\textwidth]{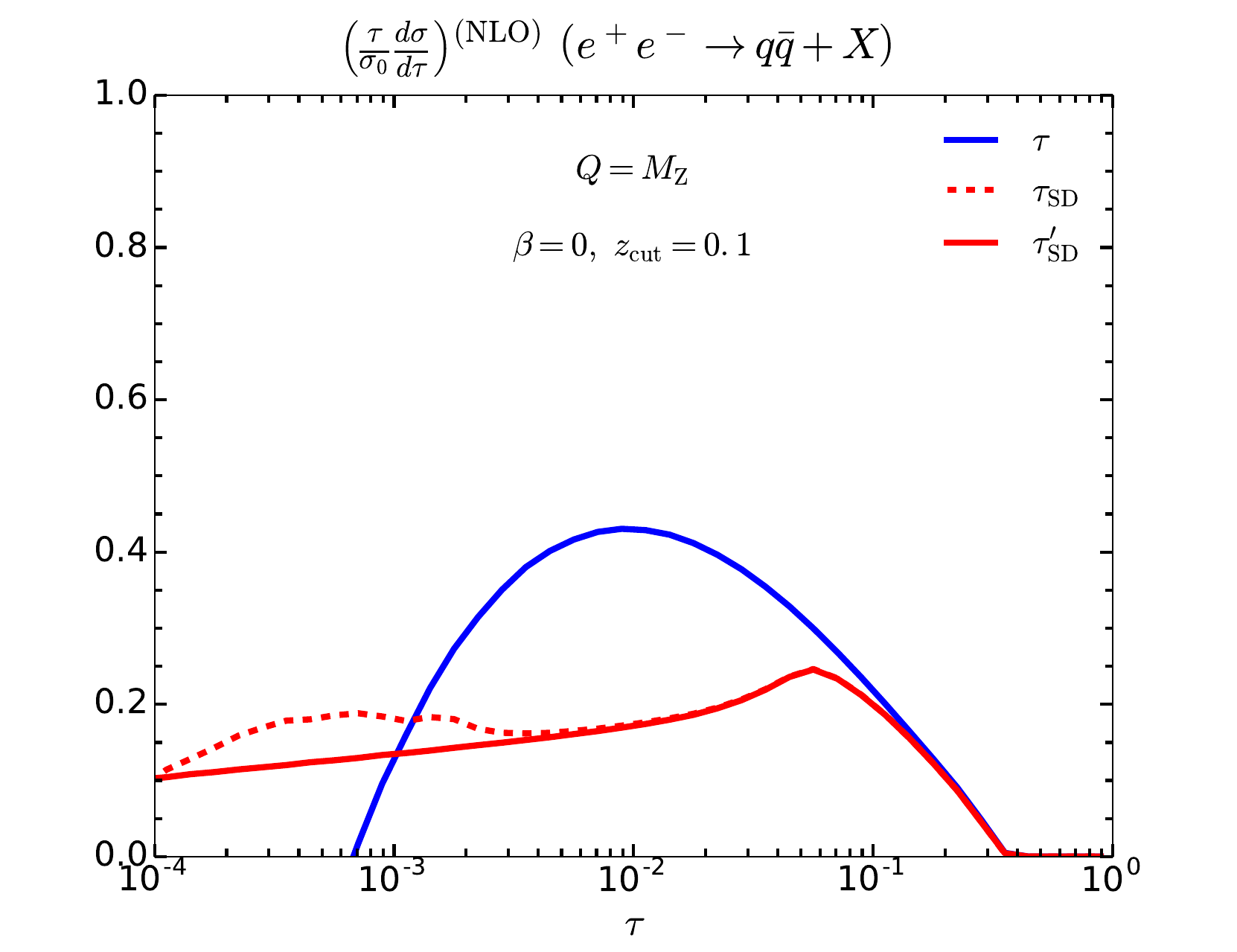}
	\caption{Fixed order thrust calculated from \eventtwo, LO on the left and NLO on the right. The dotted red lines show the original definition of thrust, while the solid red lines show its new incarnation. The two definitions coincide for the ungroomed case (solid blue), while for soft-drop thrust the new version $\tausd'$ removes the second transition region.}
	\label{Fig:Fixed-order}
\end{figure}

The soft-drop grooming technique~\cite{Larkoski:2014wba} is defined for a jet with radius $R$ using Cambridge-Aachen (C/A) clustering~\cite{Dokshitzer:1997in,Wobisch:1998wt} as:
\begin{enumerate}
	\item Undo the last step of the clustering for the jet, $J$, and split it into two subjets.
	\item Check if these subjets pass the soft drop condition, which is defined for $e^+ e^-$ collisions as~\cite{Frye:2016aiz}:
	\begin{equation}
	\frac{\min\[E_i,E_j\]}{E_i+E_j}>\zc \(\frac{1-\cos\theta_{ij}}{1-\cos R}\)^{\beta/2}
	\end{equation}
	where $E_i$ and $E_j$ are the energies of the two subjets and $\theta_{ij}$ is the angle between them. 
	\item If the splitting fails this condition the softer subjet is dropped and the groomer continues to the next step in the clustering. In other words the jet $J$ is set to be the harder of the two subjets.
	\item If the splitting passes this condition the procedure ends and the jet $J$ is the soft-drop jet.
\end{enumerate}
Soft drop has two different parameters: $\zc$, which is an energy threshold, and $\beta$, which is the angular exponent that controls how strongly wide-angle emissions are discarded. 
In the limit $\beta\to \infty$ the ungroomed jet is recovered, while $\beta=0$ corresponds to mMDT~\cite{Dasgupta:2013ihk}.
In our studies, we will heavily use jets defined by a hemisphere of the event. In this case, we find it more convenient to work with a soft-drop condition defined with a slightly different normalisation:
\begin{equation}
\frac{\min\[E_i,E_j\]}{E_i+E_j}>\zc \(1-\cos\theta_{ij}\)^{\beta/2}.
\end{equation}
The observable we will be making use of for most of this work is thrust~\cite{Farhi:1977sg}, which is defined by
\begin{equation}\label{Tdef}
T=\max_{\vec{n}}\left(\frac{\sum_{i \in \mathcal{E}}\left|\vec{n}\cdot\vec{p}_{i}\right|}{\sum_{i \in \mathcal{E}}\left|\vec{p}_{i}\right|}\right),
\end{equation}
where the $\vec{p}_{i}$ are the three-momenta of all the different particles $i$ in the event $\mathcal{E}$. The unit vector $\vec{n}$ which maximizes the sum is called the thrust axis. 
Often, especially in the context of all-order calculations, the variable
\begin{equation}\label{taudef}
\tau=1-T=\min_{\vec{n}}\left(1-\frac{\sum_{i}\left|\vec{n}\cdot\vec{p}_{i}\right|}{\sum_{i}\left|\vec{p}_{i}\right|}\right)
\end{equation}
is defined.
This observable is equal to zero for two back to back particles, however with additional emissions the observable moves away from zero.
The $\tau \ll 1$ region, often referred to as the two-jet region, is characterised by soft and collinear emissions, while
larger values of $\tau$ require hard emissions to contribute.
Given the above considerations, we are tempted to define soft-drop thrust as follows:
\begin{enumerate}[label=(\alph*)]
\item the thrust axis $n_T$ is calculated, thus dividing the event into two hemispheres;
\item the soft-drop algorithm is applied in each hemisphere;
\item \label{c} the set of particles which are left after soft drop constitutes the soft-drop event $\mathcal{E}_\textsc{SD}$, on which the soft-drop thrust $T_\textsc{SD}$ is defined as
\begin{equation}\label{TSDdef}
T_\textsc{SD}=\max_{\vec{n}} \left(\frac{\sum_{i \in \mathcal{E}_\textsc{SD}}\left|\vec{n}\cdot\vec{p}_{i}\right|}{\sum_{i\in \mathcal{E}}\left|\vec{p}_{i}\right|}\right)
+ \frac{\sum_{i\notin \mathcal{E}_\textsc{SD}}\left|\vec{p}_{i}\right|}{\sum_{i\in \mathcal{E}}\left|\vec{p}_{i}\right|},
\end{equation}
where the last term ensures IRC safety for every $\beta \ge0$, including the potentially problematic case of $\beta=0$~\cite{Marzani:2019evv}.
\end{enumerate}
Furthermore, in analogy with Eq.~(\ref{taudef}), we also introduce $\tausd=1-T_\textsc{SD}$.
The above definition seems very natural, as it is a straightforward extension of the ungroomed thrust. However step~\ref{c} does result in undesirable features. 
Let us consider for instance the $\beta=0$ case, for which soft drop coincides with mMDT. Due to the close resemblance of the $\tau$ variable with the hemisphere jet mass~\cite{Catani:1991kz,Catani:1992ua} in the soft-collinear region, we expect the $\tausd$ distribution, for $\beta=0$, to only exhibit single logarithms at small $\tausd$, which are of collinear origin.
However, this expectation is broken already at LO.
In order to see this let us consider a three-particle configuration, which at parton level is realized by allowing one emission from the quark-antiquark dipole. If this emission is soft, it is then groomed away and the groomed event is now constituted by just two partons. However, these are not aligned and therefore they provide a non-zero value of $\tausd$.
This has to happen at values of $\tausd$ which are parametrically rather small, suppressed by two powers of $\zc$. The first power of $\zc$ comes about because we are in a region where soft drop is active, while the second one arises because we are concentrating on values of $\tau$ which would have been zero in the absence of soft drop.
We note that at asymptotically small values $\tausd$, the distribution reverts to a double-logarithmic behaviour because the value of $\tausd$ is set by the kinematics of the emission which has been groomed away and it is therefore sensitive to the soft-collinear region of phase-space. 
A more detailed analysis of this type of kinematic configuration, and the resulting $\Ord\(\zc^2\)$ transition point, is performed in Appendix~\ref{sec:2nd-trans}.
This effect can be seen in Fig.~\ref{Fig:Fixed-order} for a fixed order computation at LO (on the left) and NLO (on the right) accuracy, i.e.\  with one or two emissions off the $q\bar q$ dipole calculated with the program \eventtwo~\cite{Catani:1996jh,Catani:1996vz}.
The ungroomed thrust distribution is shown in solid blue, while the naive soft-drop thrust in dotted red. The unwanted double-logarithmic behaviour of the soft-drop distribution is clearly evident. 

The resummation of the above type of contributions does not appear to be straightforward. Although these effects are confined to a rather small region of phase-space, where non-perturbative effects dominate, we find their presence a nuisance and we prefer to get rid of them altogether. 
Therefore, we modify the last step of the soft-drop thrust definition as follows:
\begin{enumerate}[label=(c$'$)]
\item \label{cp} the sets of particles left in the two hemispheres after soft drop constitute the soft-drop hemispheres $\mathcal{H}^L_\textsc{SD}$ and $\mathcal{H}^R_\textsc{SD}$, on which the soft-drop thrust $T'_\textsc{SD}$ is defined as
\begin{equation}\label{TSDpdef}
T'_\textsc{SD}=
\frac{\sum_{i \in \mathcal{H}^L_\textsc{SD}}\left|\vec{n_L}\cdot\vec{p}_{i}\right|}{\sum_{i\in \mathcal{E}}\left|\vec{p}_{i}\right|}+
\frac{\sum_{i \in \mathcal{H}^R_\textsc{SD}}\left|\vec{n_R}\cdot\vec{p}_{i}\right|}{\sum_{i\in \mathcal{E}}\left|\vec{p}_{i}\right|}
+
\frac{\sum_{i\notin \mathcal{E}_\textsc{SD}}\left|\vec{p}_{i}\right|}{\sum_{i\in \mathcal{E}}\left|\vec{p}_{i}\right|}.
\end{equation}
where $\vec{n}_L$ and $\vec{n}_R$ are the jet axes of the left and right hemispheres, respectively.~\footnote{We thank Gregory Soyez for discussions on this point. Furthermore, we note that this approach shares some similarities to event shapes defined with respect two broadening axes~\cite{Larkoski:2014uqa}.} 
If no soft drop is applied, $T'_\textsc{SD}$ reduces to $T$, as it should. Moreover, $T'_\textsc{SD}$ is free of the undesired transition point in the soft-collinear region. Again, in analogy with Eq.~(\ref{taudef}), we also introduce $\tausd'=1-T'_\textsc{SD}$.
\end{enumerate}
The LO and NLO distributions for $\tausd'$ are also shown in Fig.~\ref{Fig:Fixed-order} with solid lines. We see that the large-$\tau$ behaviour of the three distributions is identical, while $\tausd'$ has the desired behaviour in the infra-red region.
 
 \new{Finally, we note that having abandoned the use of the thrust axis in step \ref{cp}, we could also question its role in defining the hemispheres. For instance, we could directly cluster the event into two C/A jets. We have checked that the resulting distributions have very similar features to the ones obtained with the current definition.}

\section{Hadronisation corrections: a Monte Carlo study}\label{sec:MC}
\begin{figure}
	\centering
	\includegraphics[width=0.49\textwidth]{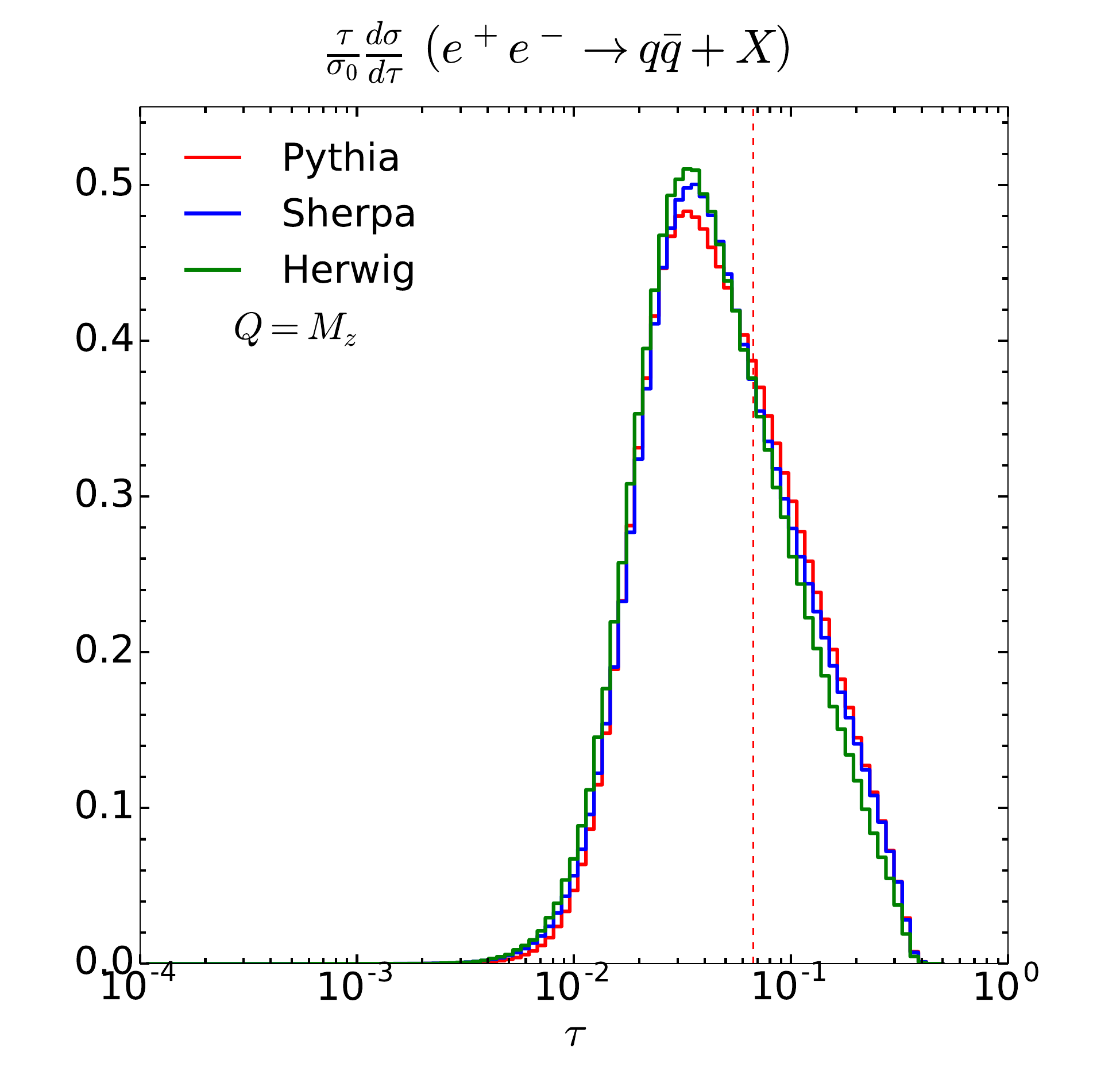}
	\includegraphics[width=0.49\textwidth]{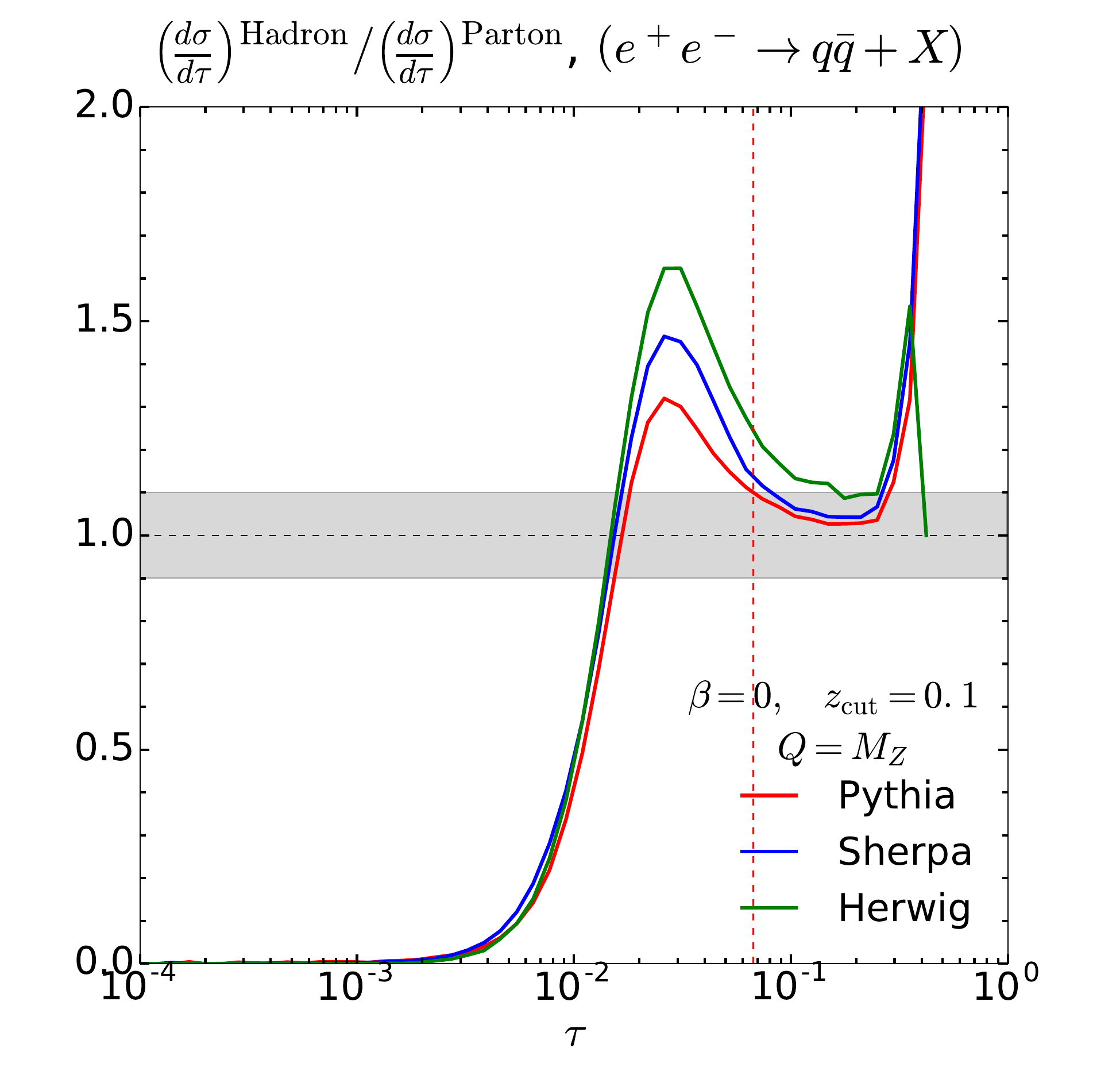}
	\includegraphics[width=0.49\textwidth]{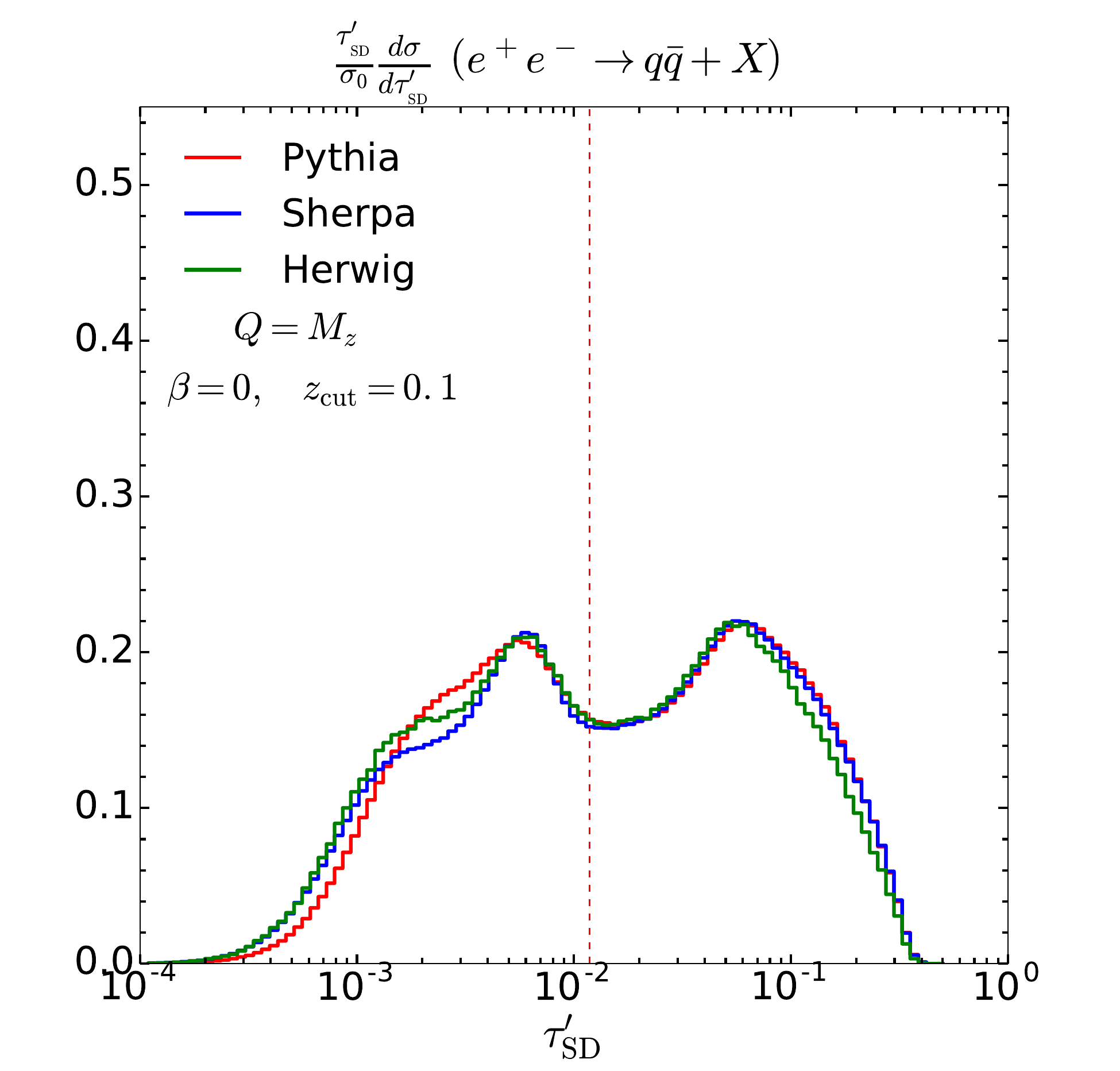}
	\includegraphics[width=0.49\textwidth]{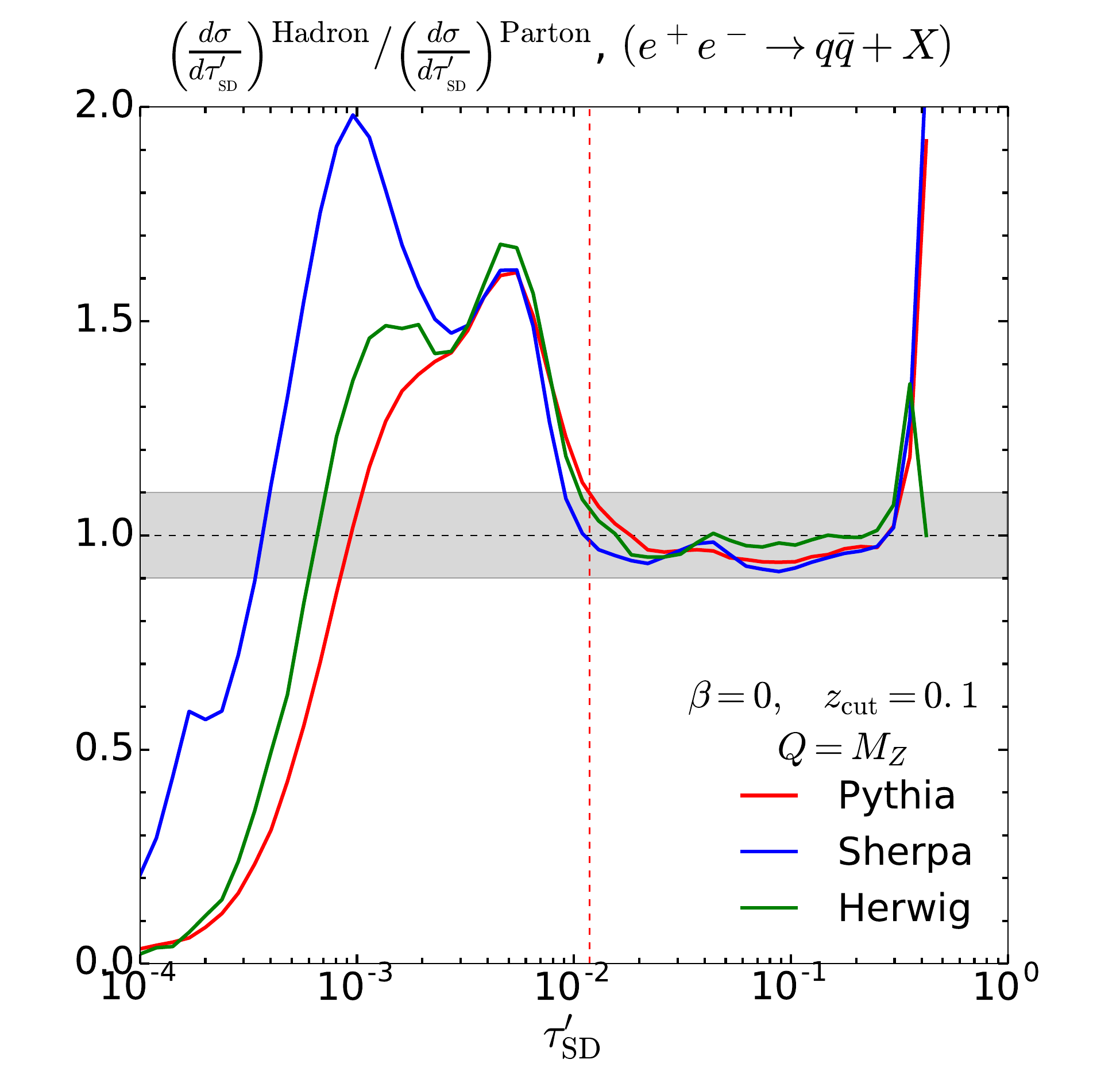}
	\caption{Monte Carlo parton shower simulation of the thrust distribution $\tau=1-T$ in $e^+e^-$ collisions, obtained using three general-purpose programs, as indicated in the legends. The plots at the top refer to the traditional (ungroomed) thrust distribution, while the bottom ones are for its soft-drop version $\tausd'=1-T_\textsc{SD}'$. The plots on the left show the normalised distribution, at the hadron level, while the plots on the right show the ratio of distributions obtained with hadronisation turned on and off. We take these ratios as proxies to assess non-perturbative corrections.
	The vertical lines correspond to the values of thrust, or soft-drop thrust, for which hadronisation corrections, as estimated using the \pythia~simulation, reach the 10\% level.}
	\label{fig:thrust_showers}
\end{figure}

The main motivation for introducing groomed event-shapes is to reduce their dependence on non-perturbative physics, such as hadronisation corrections. We note that existing studies for closely-related jet observables in proton-proton collisions, e.g.\ \cite{Marzani:2017mva,Larkoski:2014wba,Dasgupta:2013ihk,Dasgupta:2016ktv,Salam:2016yht}, are typically performed at energies of interest for LHC phenomenology, i.e.\ jets with transverse momenta of several hundred GeV. On the other hand, the bulk of the thrust data that enters current determinations of the strong coupling have been collected by the LEP experiments, i.e.\ at a centre-of-mass energy around the $Z$ mass. 
It is therefore necessary to perform a dedicated study to see if grooming techniques prove useful in reducing the size of hadronisation corrections even at the $Z$ pole.

In order to assess the size of hadronisation corrections in the thrust distribution we resort to a Monte Carlo study. Namely, we simulate $e^+e^-$ at the centre-of-mass energy $Q=m_Z$ using three different Monte Carlo parton showers: \pythia~8.219~\cite{Sjostrand:2006za,Sjostrand:2007gs}, \sherpa~2.2.3~\cite{Gleisberg:2008ta,Schumann:2007mg,Gleisberg:2008fv}, and \herwig~7.1.2~\cite{Bahr:2008pv,Bellm:2015jjp}. 
The Monte Carlo samples are generated at Born level, with default settings for the shower parameters and the hadronisation models.
Thrust and the thrust axis are computed using the implementation found in \pythia~\cite{Sjostrand:2006za}.
In order to calculate $T_\textsc{SD}'$, we use the thrust axis to partition each event into two hemispheres and we apply the $e^+ e^-$ version of the soft drop algorithm described in the previous section, making use of \fastjet~3.2.1~\cite{Cacciari:2011ma} in order to obtain the Cambridge/Aachen~\cite{Dokshitzer:1997in,Wobisch:1998wt} trees. 

The top-left panel of Fig.~\ref{fig:thrust_showers} shows the result of the three Monte Carlo simulations for the variable $\tau$, without any grooming.  We find good agreement between the three parton-shower programs, which does not come as a surprise. Indeed QCD radiation from $q \bar q$ dipoles is very well constrained by LEP data, which, in turn, are used to tune Monte Carlo parton showers. 
The main motivation for performing this numerical simulation is to assess the role of non-perturbative corrections. We address this in the top-right panel of Fig.~\ref{fig:thrust_showers}, where we show the ratio of the hadron-level simulations to their partonic counterparts, which are obtained by switching off the hadronisation process. 
We take this ratio as a reasonable proxy for the size of non-perturbative corrections~\footnote{The \herwig \ curve shows a slightly different behaviour at parton level, resulting in a visible difference in the hadron/parton ratio. Despite dedicated studies, we were not able to identify the source of this discrepancy.}. 
We note that non-perturbative corrections are sizeable for both large and small values of $\tau$. In particular, the end-point at large $\tau$ is determined by multiple resolved emissions, which are difficult to model in perturbation theory.
For this reason, the upper limit of the fitting region is sometimes taken at $\tau=1/3$, which is the end-point of the LO thrust distribution.
At the opposite end of the spectrum, i.e.\ small $\tau$, the distribution becomes sensitive to QCD at low scales and we therefore expect non-perturbative corrections to dominate. 
In the plot we mark with a vertical line the value of thrust for which hadronisation corrections, as estimated using the \pythia~simulation, reach the 10\% level. We note that this happens for $\tau\simeq 7 \cdot10^{-2}$. 
For this reason, the study of Ref.~\cite{Abbate:2010xh} also introduced a lower limit for their fitting region to ensure that perturbation theory, both fixed-order and resummed, provides the bulk of the contribution, while non-perturbative physics is a small, albeit non-negligible, correction.

In the lower-panels of Fig.~\ref{fig:thrust_showers} we show the corresponding distributions and ratios for the soft-drop version of thrust, namely $\tausd'$. We first show results for $\zc=0.1$ and $\beta=0$, which are the preferred values in LHC analyses.
Let us focus on the bottom-right plot, which shows the size of non-perturbative corrections. We see that the situation at the large-$\tausd'$ tail has not changed much compared to the ungroomed case. Indeed, we have no reason to believe that soft drop should provide any advantage in this kinematic region.
The situation is rather different at medium and small $\tausd'$. In this region, non-perturbative corrections appear to be very much reduced: indeed the ratio hadron-level to parton-level remains close to one down to smaller values of $\tausd'$, thus extending the validity of perturbation theory down to smaller values of the event shape. 
More quantitatively, the value of the observable for which hadronisation corrections reach the 10\% level is now 
$\tausd'\simeq  10^{-2}$.

\new{A wider range for the observable's values for which we trust perturbation theory is not the only criterion we need to satisfy, as we have to make sure that the soft-drop procedure does not reduce the cross-section in the desired region. 
By looking at the vertical lines on the cross-section plots (left-hand side of Fig.~\ref{fig:thrust_showers}), which indicate where non-perturbative corrections reach the 10\% level, we see that for the un-groomed case, only a third of the cross-section is in the perturbative region, while this fraction nearly doubles in the case of soft-drop thrust.
Furthermore, the ratio plots in Fig.~\ref{fig:thrust_showers} hint to another possible benefit in using soft drop, namely a reduction in the spread of the Monte Carlo estimates of hadronisation corrections. However, because we have only considered three different showers here, we cannot draw firm conclusions on this last observation from our study.}

In summary, this Monte Carlo study supports our initial intuition:  soft drop appears to be an efficient way to reduce contamination of non-perturbative physics even in $e^+e^-$ collisions at LEP energies.

\begin{figure}
\centering
\includegraphics[width=0.49\textwidth]{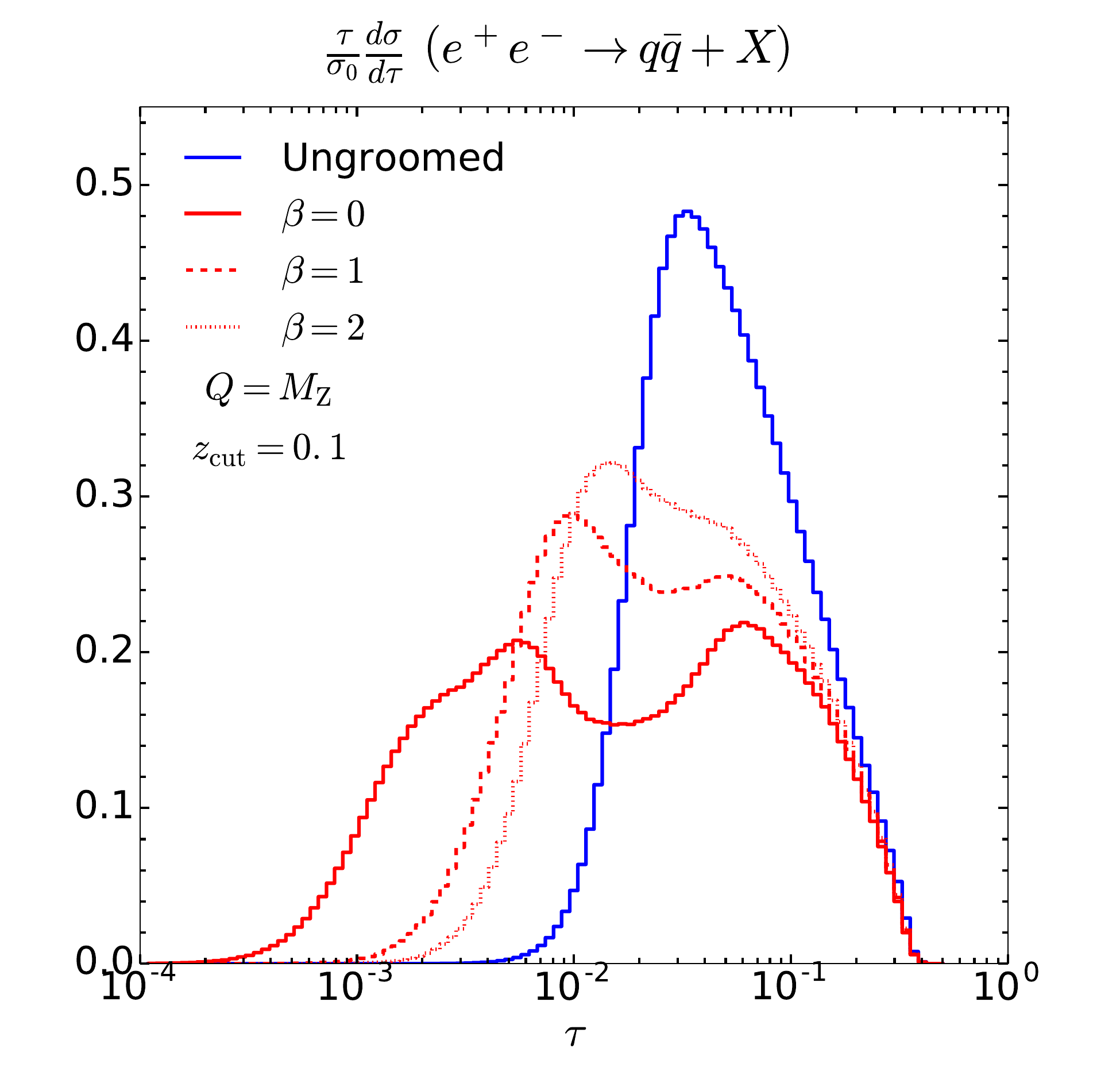}
\includegraphics[width=0.49\textwidth]{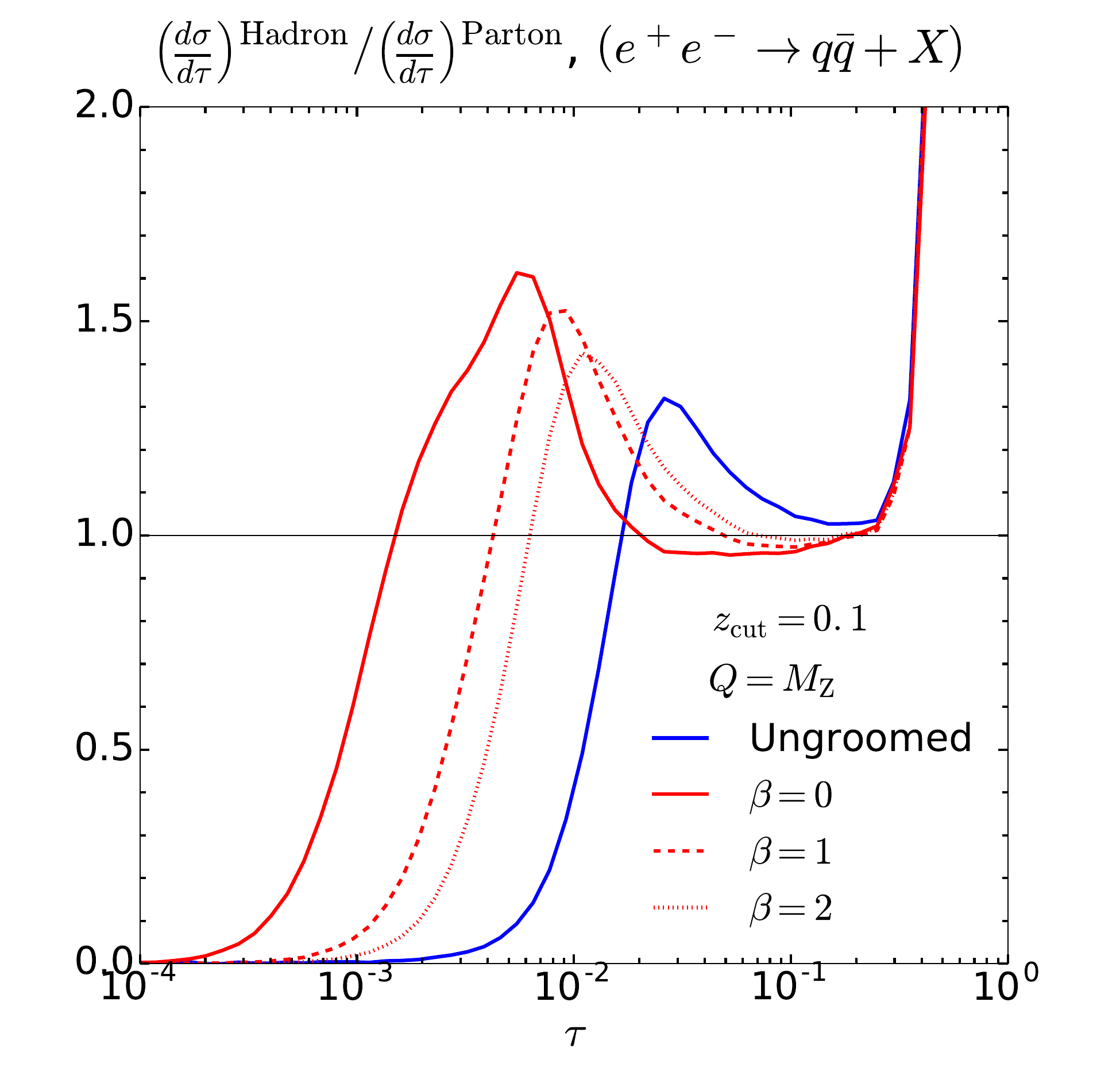}
\caption{Soft-drop thrust distribution generated with \pythia~for three different values of the angular exponent $\beta$, compared to ungroomed thrust. As $\beta$ increases we move closer to the ungroomed case, which we recover for $\beta \rightarrow \infty$. All plots are for $\zc = 0.1$.}
\label{fig:betas}
\end{figure}

\begin{figure}
\centering
\includegraphics[width=0.49\textwidth]{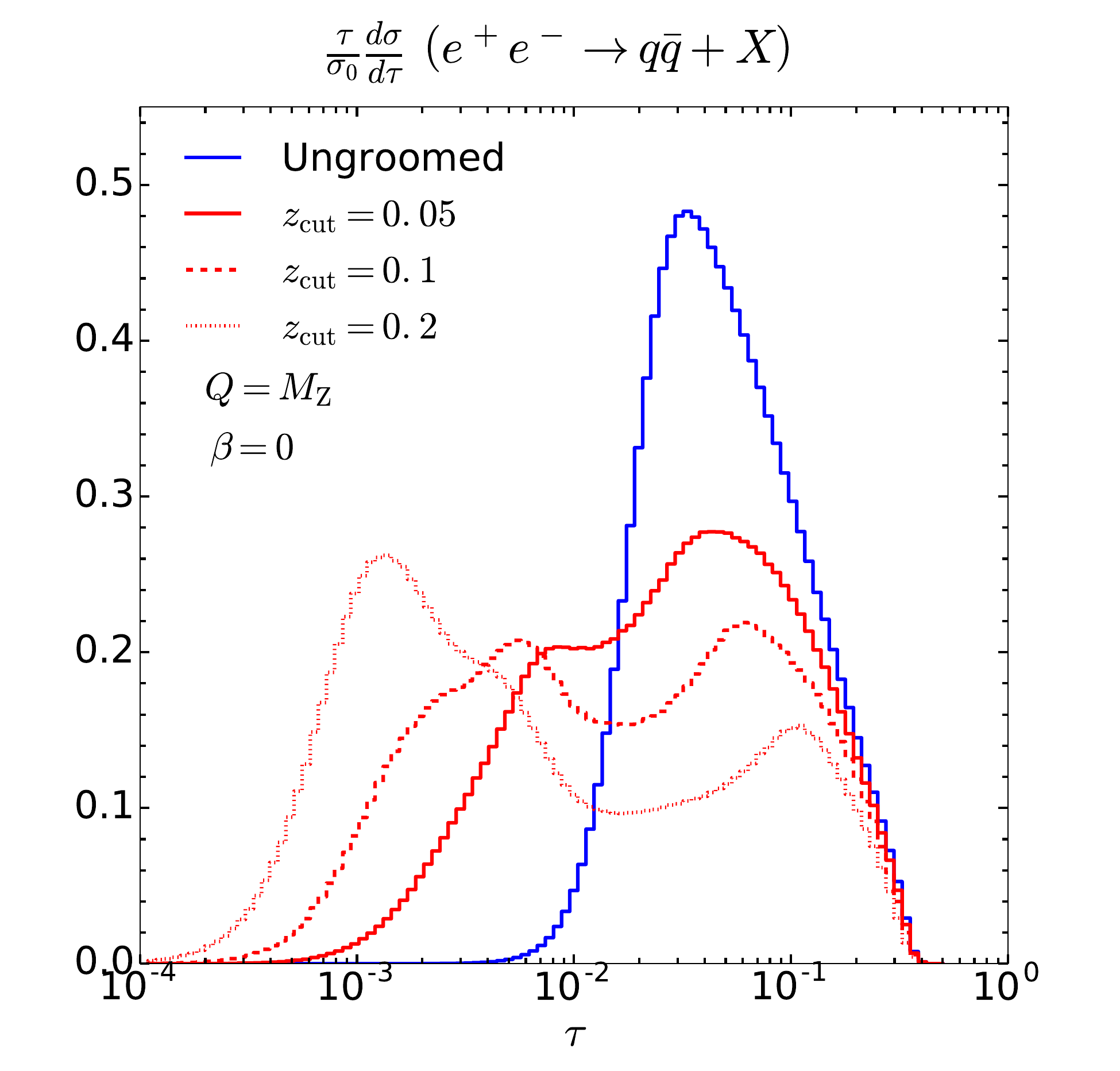}
\includegraphics[width=0.49\textwidth]{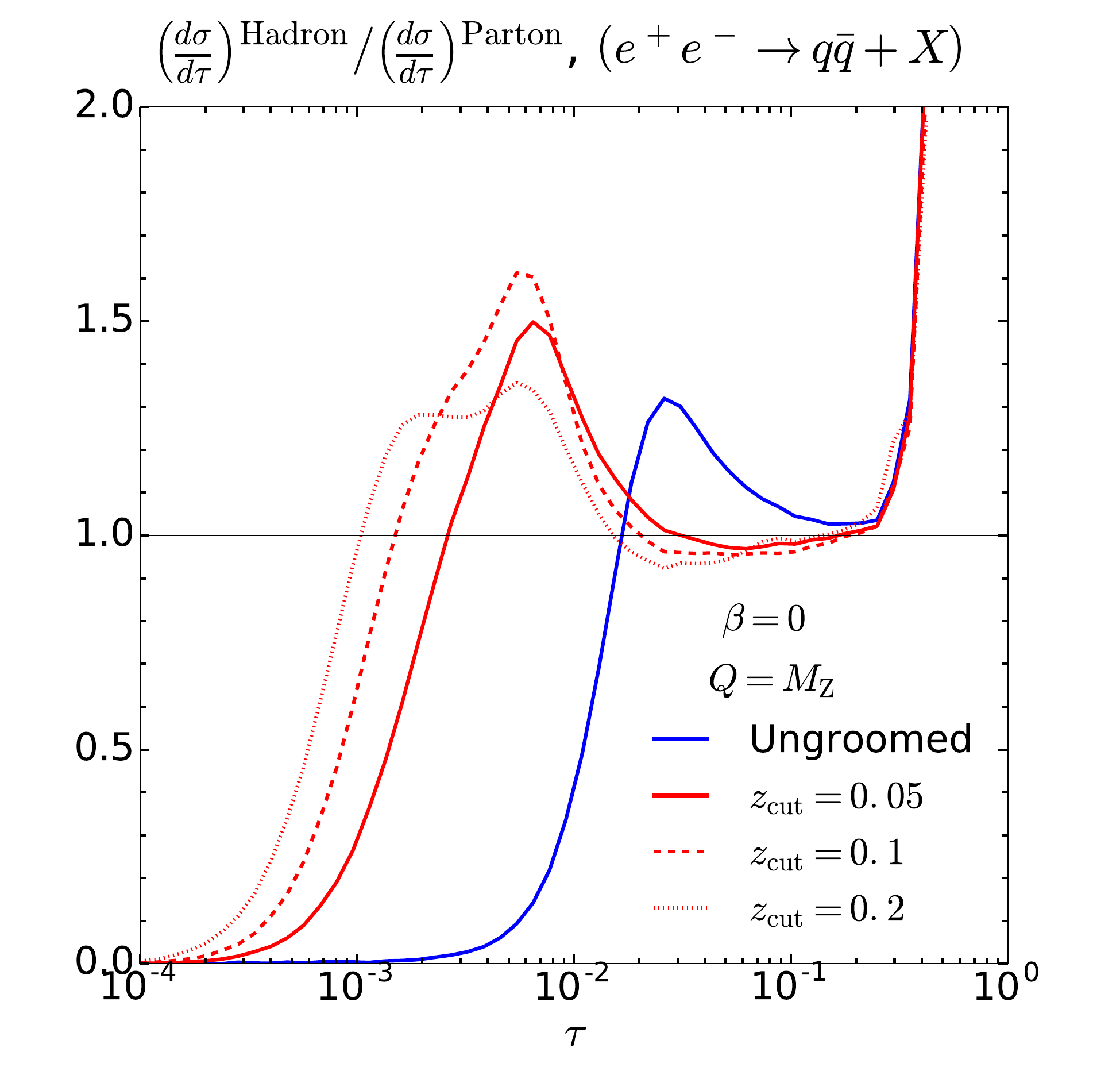}
\caption{Soft-drop thrust distribution generated with \pythia~for three different values of the energy cut $\zc$, compared to ungroomed thrust. All plots are for $\beta=0$.}
\label{fig:zcuts}
\end{figure}

Thus far we have only considered the pair of values $\zc=0.1$ and $\beta=0$, which is the preferred option for jet studies at the LHC.
However, here we are considering a different type of collision, at much lower energies, and so it is also interesting to explore other values of $\beta$ and $\zc$, in order to see if different combinations result in more desirable features. 
In particular, we would like to explore the possibility of using a milder groomer and study the corresponding trade-off in terms of sensitivity to non-perturbative physics. 
For clarity, we focus on \pythia~results, as the conclusions are similar for the other Monte Carlo generators. In Fig.~\ref{fig:betas}, we consider three different values of the angular exponent $\beta=0,1,2$, while $\zc=0.1$ is kept fixed. 
Larger values of $\beta$ correspond to milder grooming (the ungroomed result can be thought as $\beta \to \infty$).
The hadron-level results are shown as well as the ratio to the partonic level. 
Here it can be seen that the region where non-perturbative corrections are small rapidly decreases as $\beta$ increases, thus suggesting that while $\beta=1$ is still acceptable, larger values of $\beta$ do not serve our purposes.
Next, a comparison for different choices of $\zc$ can be made, which is shown in Fig.~\ref{fig:zcuts}. We choose to vary the value of $\zc$ up and down by a factor of 2, while keeping $\beta=0$.
In this case milder grooming corresponds to smaller values of $\zc$. Here we see that the choice $\zc=0.05$ is rather promising as the range over which non-perturbative corrections are small has decreased only slightly as compared to the default $\zc=0.1$, while the fraction of events in the potential $\as$ fitting region has noticeably increased.

In summary, thanks to a simple Monte Carlo study we have shown that groomed event-shapes in $e^+e^-$ collisions at LEP energies, such as soft-drop thrust, are characterised by reduced sensitivity to non-perturbative physics in the kinematic region which is typically used in determinations of the strong coupling.
Therefore, we argue that the usage of this type of observable can help with reducing the degeneracy between $\alpha_s$ and non-perturbative parameters, thus improving the reliability of strong-coupling extractions from event shapes.
However, this research program only makes sense if we are able to provide theoretical predictions for groomed event shapes which are, from a perturbative point of view, under the same theoretical control as traditional, i.e.\ ungroomed, event shapes. 
In the next section we discuss perturbative predictions for the variable $\tausd'$, emphasising its strengths and limitations.

\section{Resummation of soft-drop thrust}\label{Sec:Analytics}

The all-order resummation of the soft-drop thrust ($\tausd'$) distribution closely follows the calculation of the soft-drop energy correlation $e_2^{(2)}$, which was performed to NNLL accuracy in Ref.~\cite{Frye:2016aiz}. The result is based on the factorisation of the differential distribution in terms of hard, soft and collinear functions, which was derived using SCET. Because $e_2^{(2)}$ and thrust are proportional in the soft limit, an analogous factorisation theorem holds for soft-drop thrust. Therefore, in the $\tau \ll \zc\ll1$ limit we have
\begin{equation} \label{eq:fact_theorem}
\frac{d\sigma}{d\tau}=H\(Q\)S_{G}\(\zc,\beta\)\[S_C\(\tau,\zc,\beta\)\otimes J\(\tau\)\]^2,
\end{equation}
where in order not to clutter notation we henceforth write $\tau$ instead of $\tausd'$.
Here $H$ is the hard function, which takes into account the virtual contributions and depends only on the energy scale $Q$. The global soft function, $S_G$, takes into account soft wide-angle emissions. Since soft wide-angle radiation is groomed away by soft drop and it does not influence the value of thrust, its scale only depends on the soft-drop parameters $\zc$ and $\beta$, and on the hard scale $Q$. Collinear hard emissions are taken into account by the jet function $J$. Since collinear hard radiation always passes the soft drop condition, this function will only depend on $\tau$. Finally $S_C$ describes soft collinear emissions, i.e.\ radiation which can be groomed away but can also pass and lead to a non-zero value of thrust.  Therefore, it depends on both the groomer's parameters and on the observable. In the context of SCET, resummation of large logarithmic corrections is obtained by evolving these functions using renormalisation group equations. Therefore, it is fully determined by the knowledge of the fixed-order expansions of the above functions and their anomalous dimensions. For a generic function $K$, we have 
\begin{equation}
\mu \frac{d K(\mu)}{d\mu }= \left[\Gamma_K(\as) \log \frac{\mu^2}{\mu_K^2} + \gamma_K(\as)\right] K(\mu),
\label{eq:ren-eq}
\end{equation}
where the terms in the square bracket are, respectively, the cusp and non-cusp contributions to the anomalous dimension, while $\mu_K$ is the infra-red scale of the function we are considering. 
The above factorisation theorem is valid in the asymptotic limit $\tau \ll \zc$ and, as usual, it holds up to power-corrections in the observable. 
We anticipate that, together with this asymptotic, we are also going to explore the region $\tau \sim \zc$ in detail, discussing how power-corrections to the factorisation theorem become order-one contributions in this kinematic region. 
The main results are summarised in the following, while more details are collected in Appendix~\ref{app:details}. 
%
%
\new{Here, we only explicitly discuss the relevant functions at one loop, which are needed for NLL$'$ resummation.}

The hard function $H$ is determined by the virtual corrections to the $e^+e^-\to q\bar q$ process. These depend on the underlying process but not on the observable nor on the grooming algorithm.  At one loop, the result is~\cite{Bauer:2003di,Manohar:2003vb,Ellis:2010rwa,Bauer:2011uc}
\begin{equation}
H=1+\frac{\as}{2\pi}C_F \(\frac{\mu^2}{Q^2}\)^{\epsilon}\[-\frac{2}{\epsilon^2}-\frac{3}{\epsilon}+\frac{7\pi^2}{6}-8\] +\mathcal{O}\left( \as^2 \right),
\label{eq:Hard}
\end{equation}
where we have absorbed the $\MSbar$ constant in the definition of the scale $\mu$.
The hard scale is given by $\mu_H=Q$.
The coefficient on the double pole determines the cusp contribution
\begin{equation}
\Gamma_H = -2 C_F \Gamma_{\rm cusp},
\end{equation}
where $\Gamma_{\rm cusp}$ is the universal cusp anomalous dimension~\cite{Korchemsky:1987wg}
\begin{equation}
\Gamma_{\rm cusp}=\sum_{n=0}^{\infty}\Gamma_n \(\frac{\as}{\pi}\)^{n+1}, \quad \text{with} \quad
\Gamma_0=1,\; \Gamma_1=\frac{C_A}{2} \(\frac{67}{18}-\frac{\pi^2}{6}\)-\frac{5}{9}T_R n_f.
\end{equation}
The non-cusp anomalous dimension can be written as
\begin{equation}
\gamma_H=\sum_{n=0}^{\infty}\gamma^{(n)}_H\(\frac{\as}{\pi}\)^{n+1},
\end{equation}
where the one-loop coefficient can also be found from Eq.~(\ref{eq:Hard}) based on the coefficient for the single pole:
\begin{equation}
\gamma^{(0)}_{H}=-3 C_F.
\end{equation}

The jet function is obtained by considering emissions in the collinear limit. 
The calculation is observable-dependent, however it is not groomer-dependent. Therefore the same results as for un-groomed thrust can be used here~\cite{Bosch:2004th}. Furthermore, in order to diagonalize the convolution product in Eq.~(\ref{eq:fact_theorem}) we consider Laplace moments
\begin{align}\label{eq:jet-fnct-res}
\tilde{J}\(N\)&=\int_{0}^{\infty}d\tau e^{-N\tau} J\(\tau\)
=C_F\frac{\as}{2\pi}\[1+\frac{\pi^2}{12}\frac{\partial^2}{\(\partial\log N\)^2}\]\(\frac{\bar N\mu^2}{Q^2}\)^{\epsilon}\[\frac{2}{\epsilon^2}+\frac{3}{2\epsilon}+\frac{1}{2}\(7-\pi^2 \) \],
\end{align}
where $\bar N= N e^{\gamma_E }$.
From this result, we conclude that the collinear scale is given by $\mu_J^2=Q^2/\bar{N}$. Furthermore, the coefficient of the double and single poles determine the cusp and non-cusp contributions to the jet-function anomalous dimension
\begin{align}
\Gamma_J &= 2 C_F \Gamma_{\rm cusp},\\
\gamma^{(0)}_{J}&=\frac{3}{2}C_F.
\end{align}

Finally, we turn our attention to the two soft functions that enter the factorisation theorem. The global soft function does not depend on the observable, but only on the soft-drop parameters~\cite{Frye:2016aiz}. Its expression reads
\begin{align} \label{eq:global-soft-res}
S_G\(\zc,\beta\)&=
\frac{\as C_F}{2\pi}\(\frac{\mu}{2^{\beta/2} \zc  Q}\)^{2\epsilon}\[\frac{2}{\beta+1}\frac{1}{\epsilon^2}-\frac{\pi^2}{6}\(\frac{1}{1+\beta}+2+\beta\)\],
\end{align}
from which we can easily deduce that the global soft scale is $\mu_{S_G}=2^{\beta/2} \zc Q$, while the anomalous dimensions are
\begin{align}
\Gamma_{S_G} &= \frac{2}{\beta+1} C_F \Gamma_{\rm cusp},\\
\gamma^{(0)}_{S_G}&=0.
\end{align}
The collinear soft function is both groomer and observable dependent. Its Laplace space expression is
\begin{align} \label{eq:coll-soft-res}
\tilde{S}_C\(N,\zc,\beta\)&=
\frac{\as}{2\pi}C_F\[1+\frac{\pi^2}{12}\frac{\partial^2}{\(\partial\log N\)^2}\]\(\frac{\mu \bar{N}^{\frac{\beta+1}{\beta+2}}}{2^{\frac{\beta/2}{\beta+2}} \zc^{\frac{1}{\beta+2}}  Q}\)^{2\epsilon}\frac{\beta+2}{\beta+1}\[-\frac{1}{\epsilon^2}+\frac{\pi^2}{12}\],
\end{align}
from this the soft-collinear scale can be read off as $\mu_{S_C}=\[\frac{2^{\beta/2} \zc}{\bar{N}^{\beta+1}}\]^{\frac{1}{\beta+2}} Q$, while the anomalous dimensions are
\begin{align}
\Gamma_{S_C} &=-\frac{\beta+2}{\beta+1} C_F \Gamma_{\rm cusp},\\
\gamma^{(0)}_{S_C}&=0.
\end{align}

Cancellation of all singularities and renormalisation-group invariance at the order considered here require that
\begin{eqnarray}
\Gamma_{H}+\Gamma_{S_G}+2\Gamma_{S_C}+2\Gamma_{J}&=&0\label{eq:double-pole},\\
\gamma^{(0)}_{H}+\gamma^{(0)}_{S_G}+2\gamma^{(0)}_{S_C}+2\gamma^{(0)}_{J}&=&0\label{eq:single-pole},\\
\Gamma_{H}p^{(i)}_{H}+\Gamma_{S_G}p^{(i)}_{S_G}+2\Gamma_{S_C}p^{(i)}_{S_C}+2\Gamma_{J}p^{(i)}_{J}&=&0\label{eq:power-sum},
\end{eqnarray}
where $p^{(i)}_{K}$ is the power at which a variable $i=\zc, \bar N, \dots$ enters the definition of the scale $\mu_K$. 
It is easy to verify that the above relations are indeed satisfied. 
Furthermore, one can exploit the above relations, which are valid order by order in the strong coupling, to infer the two-loop anomalous dimension of the collinear soft function, the only one which is both observable and soft-drop dependent, from the knowledge of the anomalous dimension of the hard, global soft and collinear functions, which can be taken from the literature. 
This \new{is} enough to extend the resummation of soft-drop thrust to NNLL accuracy.

\subsection{The soft function in the transition region} \label{sec:trans-region}

The factorisation of the soft function in terms of the global and collinear soft functions, Eqs.~(\ref{eq:global-soft-res}) and~(\ref{eq:coll-soft-res}) respectively, is obtained in the limit $\tau\ll\zc$ or, equivalently in Laplace space $N\gg 1/ \zc$.
The region $\tau \sim \zc$ is often referred to as the transition region~\cite{Dasgupta:2013ihk, Dasgupta:2013via} because if $\tau \gsim \zc$ soft-drop is not active, and hence the soft-drop thrust is perturbatively equivalent to its ungroomed counterpart, while if $\tau \lsim \zc$ the grooming procedure does modify the emission phase-space.
One can take different approaches to the treatment of the transition region. For instance, in Ref.~\cite{Frye:2016aiz} the region $\tau \sim \zc$ was calculated at fixed-order through matching. In contrast, the jet mass study of Refs.~\cite{Marzani:2017mva,Marzani:2017kqd} did supplement the theoretical prediction in the transition region with a resummation of the ungroomed jet mass.

In this study we want to have a closer look at the transition region. Therefore, we calculate soft corrections without assuming the hierarchy $\tau\ll \zc$. At one loop, we consider the emission of a soft gluon with momentum $k$ off a dipole with light-like momenta $n$ and $\bar n$: 
\begin{align}\label{eq:soft-func-start}
{S}\(\tau, \zc, \beta\)&=g_s^2C_F\(\frac{\mu^2e^{\gammae}}{4\pi}\)^{\epsilon}\int\frac{d^{4-2\epsilon}k}{\(2\pi\)^{3-2\epsilon}}\frac{n\cdot \bar{n}}{k^- k^+}\delta\(k^2\)\Theta\(k^-+k^+\)\Theta\(k^--k^+\)\\
&\times\[\Theta\(\zc Q\[\frac{k^+}{k^0}\]^{\beta/2}-2k^0\)\delta\(\tau\)+\Theta\(2k^0-\zc Q\[\frac{k^+}{k^0}\]^{\beta/2}\)\delta\(\tau-\frac{k^+}{Q}\)\]+\(n\leftrightarrow\bar{n}\)\nonumber,
\end{align}
where $\Theta$ is the Heaviside step function.  The first term in square brackets accounts for the emission failing soft drop, while the second one for passing it. Performing the integrals is a straightforward exercise, where most clarity is offered by looking at the cumulative soft function:
\begin{align}\label{eq:soft-func-cntd}
{\Sigma}^\text{soft}\(\tau, \zc, \beta\)&=\int_0^\tau d\tau^\prime {S}\(\tau^\prime, \zc, \beta\)=\frac{\as C_F}{2\pi}\(\frac{\mu}{2^{\beta/2} \zc  Q}\)^{2\epsilon}\[\frac{2}{\beta+1}\frac{1}{\epsilon^2}-\frac{\pi^2}{6}\(\frac{1}{1+\beta}+2+\beta\)\]\nonumber\\
&+\frac{\as}{\pi}C_F\(\frac{\mu}{2^{\frac{\beta/2}{\beta+2}} \zc^{\frac{1}{\beta+2}}\tau^{\frac{\beta+1}{\beta+2}}  Q}\)^{2\epsilon}\frac{\beta+2}{\beta+1}\[-\frac{1}{\epsilon^2}+\frac{\pi^2}{12}\]
\nonumber\\&+\frac{\as C_F}{2\pi}\(2\(\beta+2\) {\rm Li}_2\[\frac{1}{2}\(\frac{2\tau}{\zc}\)^{\frac{2}{\beta+2}}\]\),
\end{align}
for $\tau\leq \zc/2$.
The dilogarithmic contribution, although power-suppressed at small-$\tau$, is crucial in order to recover the plain thrust soft function at the transition point $\zc=2\tau$:
\begin{equation}
{\Sigma}^\text{soft}\(\tau, 2\tau, \beta\)= \frac{\as C_F}{2 \pi} \left[-\frac{2}{\epsilon^2}+\frac{4 \log \tau}{\epsilon} -4 \log^2 \tau +\frac{\pi^2}{6}+\mathcal{O}(\epsilon) \right],
\end{equation}
where we have set $\mu=Q$.
Furthermore, in Laplace space, we have
\begin{align}\label{eq:soft-func-res-laplace}
\tilde{S}\(N, \zc, \beta\)&=\int d\tau\; e^{-N\tau} S\(\tau, \zc, \beta\)=
\frac{\as C_F}{2\pi}\(\frac{\mu}{2^{\beta/2} \zc  Q}\)^{2\epsilon}\[\frac{2}{\beta+1}\frac{1}{\epsilon^2}-\frac{\pi^2}{6}\(\frac{1}{1+\beta}+2+\beta\)\]\nonumber\\
&+\frac{\as}{\pi}C_F\[1+\frac{\pi^2}{12}\frac{\partial^2}{\(\partial\log N\)^2}\]\(\frac{\mu \bar{N}^{\frac{\beta+1}{\beta+2}}}{2^{\frac{\beta/2}{\beta+2}} \zc^{\frac{1}{\beta+2}}  Q}\)^{2\epsilon}\frac{\beta+2}{\beta+1}\[-\frac{1}{\epsilon^2}+\frac{\pi^2}{12}\]\nonumber\\
&+\frac{\as C_F}{2\pi}\(2\(\beta+2\) {\rm Li}_2\[\frac{1}{2}\(\frac{1}{2}\zc \bar{N}\)^{\frac{-2}{\beta+2}}\]\).
\end{align}
We note that the above soft function contains two different scales: the soft wide-angle scale $\mu_{S_G}=2^{\beta/2} \zc Q$ and the soft collinear scale $\mu_{S_C}=\[\frac{2^{\beta/2} \zc}{\bar{N}^{\beta+1}}\]^{\frac{1}{\beta+2}} Q$, which were previously defined.
In particular, the dilogarithm depends on the ratio of these two scales. 
In the limit $N \gg 1/\zc$, the contribution from the last line vanishes and the soft functions splits in the two single-scale soft functions $S_G$ and $S_C$ previously analysed. 

\subsection{Implementation of the resummation}\label{sec:implementation}

We are now ready to assemble together the results presented earlier and obtain a resummed expression for the soft-drop thrust distribution. 
We find it most convenient to present the resummed results for the cumulative distribution defined as
\begin{equation}\label{cumulative_def}
\Sigma(\tau)= \frac{1}{\sigma_0}\int_0^\tau d \tau'\frac{d \sigma}{d \tau'},
\end{equation}
where $\sigma_0$ indicates the Born cross-section.
We start by considering the logarithmic behaviour of the cumulative \new{distribution} when soft-drop is active, i.e. below the transition point $2\tau< \zc$. In the region $\tau \ll \zc$ the usual logarithmic counting holds and we can write
\begin{align}\label{eq:Integrated-xsec-N}
\Sigma\(\tau\)&=\frac{1}{2\pi i}\int_C \frac{dN}{N} \[1+\sum_{n=1}^{\infty}\(\frac{\as}{\pi}\)^{n}\tilde{C}^{\(n\)},\]e^{\tilde{F}\(\lambda_{\bar{N}},\lambda_{\zc}\)},
\end{align}
where $\tilde{C}$ encapsulates the constant contributions in $\tau$ and $\zc$ and $\as$ is computed at a scale $\mu$. The resummed exponent $\tilde{F}$ is given by
\begin{equation}
\tilde{F}\(\lambda_{\bar{N}},\lambda_{\zc}\) = \frac{1}{\as}f_1\(\lambda_{\bar{N}},\lambda_{\zc}\)+f_2\(\lambda_{\bar{N}},\lambda_{\zc}\)+\as f_3\(\lambda_{\bar{N}},\lambda_{\zc}\),
\end{equation}
where $\lambda_{x}=\as b_0 \log x$ and the functions $f_i$ take into account N$^{i-1}$LL contributions.
Furthermore, to any fixed-logarithmic accuracy, the inverse Laplace transform can be performed analytically~\cite{Catani:1992ua}.\
The resulting expression in physical ($\tau$) space has a form that closely resembles Eq.~(\ref{eq:Integrated-xsec-N})
 \begin{align}\label{eq:Integrated-xsec-tau}
\Sigma\(\tau\)&=\[1+\sum_{i=1}^{\infty}\(\frac{\as}{\pi}\)^{n}C^{\(n\)}\] \exp\[\frac{1}{\as}g_1\(-\lambda_\tau,\lambda_{\zc}\)+g_2\(-\lambda_\tau,\lambda_{\zc}\)+\as g_3\(-\lambda_\tau,\lambda_{\zc}\)\],
\end{align}
where the functions $g_i$ only depend on the functions $f_i$ and their derivatives.
Explicit expressions, as well as detailed derivations are collected in Appendix~\ref{app:details}.
Beyond the transition point $2\tau> \zc$, we instead employ the standard resummation for thrust.

However, as previously discussed, if we want to obtain a smooth transition between the groomed and ungroomed regime, we have to supplement the calculation with those contributions which are power-suppressed at small $\tau$ but $\mathcal{O}(1)$ in the transition region. 
In order to find the contribution to the resummed exponent, we therefore consider running coupling corrections to transition-region corrections:
\begin{equation}\label{trans-rc}
g_{tr}\(\tau,\zc\)=\frac{\as}{\pi}C_F\(\beta+2\) {\rm Li}_2\[\frac{1}{2}\(\frac{2 \tau}{\zc} \)^{\frac{2}{\beta+2}}\]\frac{1}{1-2\lambda_{\zc}}.
\end{equation}
The logarithmic accuracy of the above contribution is difficult to assess because in the asymptotic region of small $\tau$, it is a power correction.
To lowest order in the strong coupling, this term resembles a contribution to the overall constant $\tilde{C}^{(1)}$, however, it does receive logarithmic running coupling corrections, which are accounted for in $\lambda_{\zc}$. 
This behaviour further complicates when we consider the differential thrust distribution, because the derivative with respect to $\tau$ acts both on the dilogarithm as well as on the logarithmic part.
This contribution is able to fully resolve the discontinuity issue at the lowest order in the strong coupling. However, further issues related to multiple emissions, similar to those discussed also in~\cite{Larkoski:2014wba}, appear at $\mathcal{O}(\as^2)$ and beyond. 
Unfortunately, the method investigated in this work is not able to solve them as they are beyond the accuracy considered here.

\subsection{Numerical results for $\beta=0$ \new{and $\beta=1$}}

Now that our setup has been established, numerical results can be presented and discussed. We present results at the centre-of-mass energy $Q=m_Z$, with $\as\(m_Z\)=0.1181$~\cite{Patrignani:2016xqp} using NLO running and the five-flavour scheme. 
For this proof-of-concept study we begin by considering  NLL$^\prime$, i.e.\ we include the functions $g_1$, $g_2$ and $C^{\(1\)}$. In addition we  include the dilogarithm contribution in $C^{\(1\)}$.
Furthermore, in order to obtain a reliable description in the whole $\tau$ range, we match the resummation to fixed-order, using the program \texttt{EVENT2}~\cite{Catani:1996jh,Catani:1996vz}. We \new{first} consider tree-level (LO) matrix elements and we employ a standard additive matching:
\begin{equation}
\tau \frac{d\sigma^{\rm LO+NLL^\prime}}{d\tau}=\tau \frac{d\sigma^{\rm LO}}{d\tau}+\[\tau \frac{d\sigma^{\rm NLL^\prime}}{d\tau}-\tau \frac{d\sigma^{\rm NLL^\prime|_{LO}}}{d\tau}\],
\end{equation}  
where the last contribution subtracts the expansion of the resummation to first order,  in order to avoid double counting. 
\new{Extension to higher accuracy, both at fixed-order and resummed level is discussed in section~\ref{sec:precision}.}

\begin{figure}
	\centering
	\includegraphics[width=0.45\textwidth]{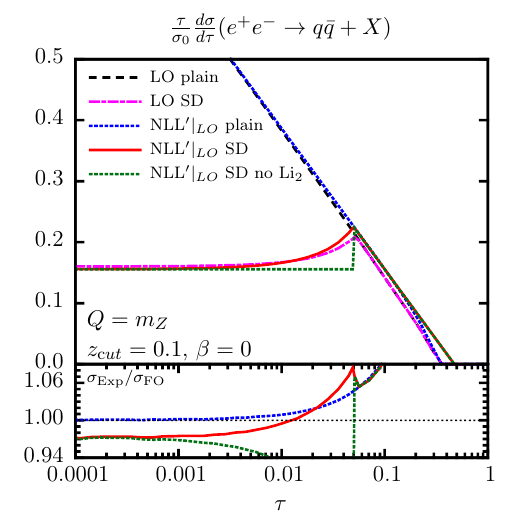}
	\includegraphics[width=0.45\textwidth]{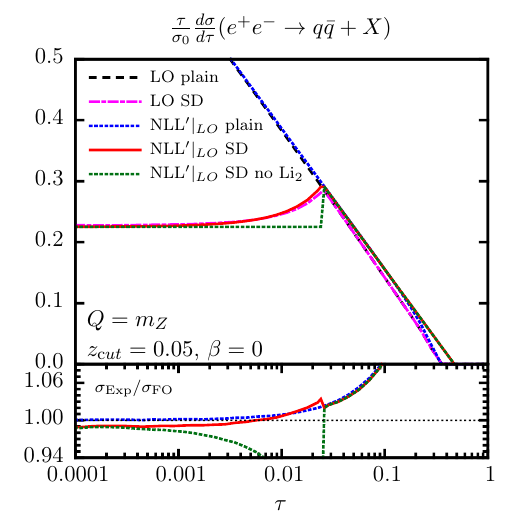}
	\caption{A comparison between the fixed order and expansion of the differential cross section of the plain and soft-dropped thrust for $e^+ e^-$ collisions at a centre of mass energy $Q=m_Z$ and the soft drop parameters $\beta=0$ and $\zc=0.1$ (left) and $\zc=0.05$ (right). The figure shows LO for plain thrust (black dashed) and soft-dropped thrust (magenta dashed-dotted) and the expansion for plain thrust (blue dotted) and soft-dropped thrust (red solid).	 \new{The plots at the bottom show the ratio of the expansions to their fixed-order counterparts.}} 
	\label{fig:exp-SD-tau}
\end{figure}

\new{We start by considering the $\beta=0$ case}~\footnote{It is well-known~\cite{Dasgupta:2013ihk,Larkoski:2014wba} that the logarithmic counting changes in this case, as logarithms of soft origin disappear. Therefore, NLL terms become the first non-vanishing contribution and they could be referred to as "LL". However, for consistency with the rest of the paper, we prefer to keep here the counting for general $\beta$.}
In Fig.~\ref{fig:exp-SD-tau}, we compare the fixed-order calculation for plain thrust and the soft-drop thrust to the first-order expansion of the resummation. We do this for two different values of $\zc$: $\zc=0.1$ on the left and $\zc=0.05$ on the right.

\new{The ratio plots at the bottom show} that the expansion of the resummation correctly captures the asymptotic behaviour of the LO distribution in all cases. We have verified that the slight off-set in the soft-drop distributions is due to power corrections in $\zc$, which we do not account for here (see~\cite{Dasgupta:2013ihk,Marzani:2017mva} for studies of their impact on similar observables).
Furthermore, we note that the solid red curve is continuous in the transition region, because of the dilogarithmic correction terms that we have introduced, while the discontinuity is clearly visible if this term is dropped, as shown in the green dotted curve. 

\begin{figure}
	\centering
	\includegraphics[width=0.45\textwidth]{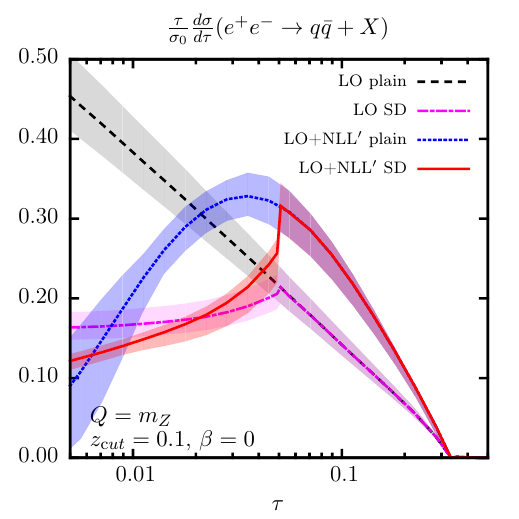}
	\includegraphics[width=0.45\textwidth]{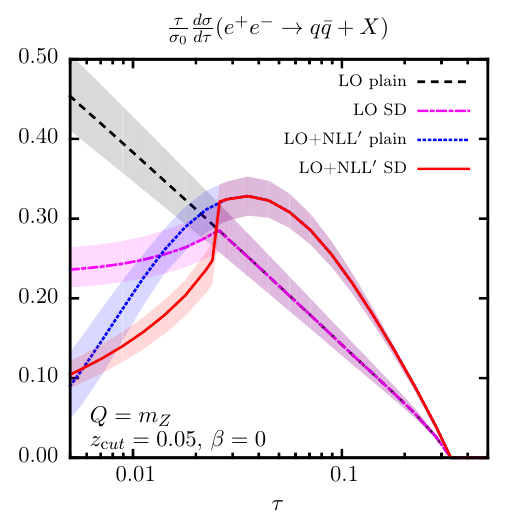}
	\caption{The results for the fixed order and resummed differential cross section of the plain and soft-dropped thrust for $e^+ e^-$ collisions at a centre of mass energy $Q=m_Z$ and the soft drop parameters $\beta=0$ and $\zc=0.1$ (left) and $\zc=0.05$ (right). The figure shows LO for plain thrust (black dashed) and soft-dropped thrust (magenta dashed-dotted) and the matched LO+NLL$^\prime$ cross section for plain thrust (blue dotted) and soft-dropped thrust (red solid) with the bands from resummation and scale uncertainties included.} 
	\label{fig:matched-SD-tau}
\end{figure}

We then present fully resummed and matched results in Fig.~\ref{fig:matched-SD-tau}. In order to ease the matching procedure, we have modified the argument of the logarithms, so that the end-point of the resummed distribution matches the one of the fixed-order $\tau_{\rm max}=1/3$~\cite{Catani:1992ua}. The details of this prescription are given in Appendix~\ref{app:resum_formulae}.
We estimate the perturbative uncertainty by \new{perfoming a 7-point scale variation of renormalisation and resummation scales around their central value $Q$, i.e.\ we vary both scales by a factor of two up and down but we discard combinations that give rise to logarithms of four.}
We first note that in the case of plain thrust, the impact of resummation is significant, which shows that higher-order corrections are sizeable.
As expected, this remains true for soft-drop thrust for values of $\tau$ above the transition region.
However, we note that if $\zc$ is not too small, so that logarithms of $\zc$ do not play an important role, then resummation becomes a less significant corrections to the fixed-order calculation. An observation which was already made in the context of high-$p_t$ jet mass distributions after grooming~\cite{Dasgupta:2013ihk,Marzani:2017mva,Marzani:2017kqd}, which remains true also in this context.
Furthermore, we remind the reader that from the analysis performed in section~\ref{sec:MC}, hadronisation corrections become sizeable, i.e.\ bigger than 10\%, for values of the ungroomed $\tau$ below $7 \cdot 10^{-2}$, while if soft-drop is applied this happens for $\tau \lesssim   10^{-2}$.

\begin{figure}
	\centering
	\includegraphics[width=0.45\textwidth]{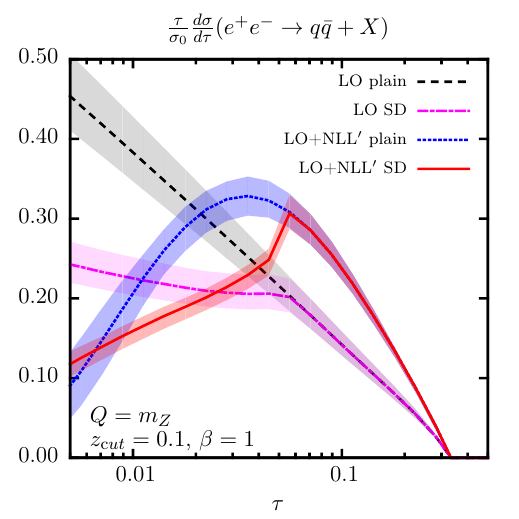}
		\includegraphics[width=0.45\textwidth]{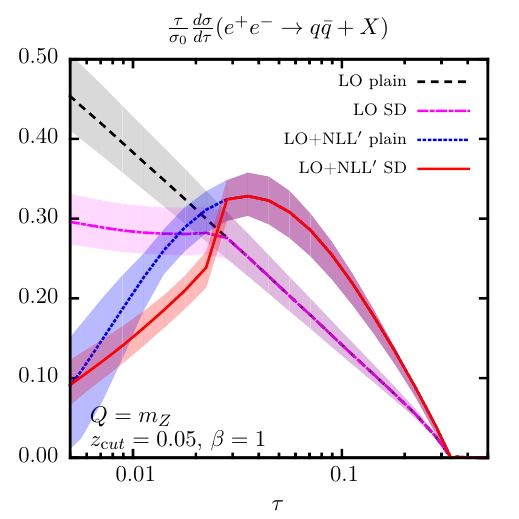}
	\caption{\new{Same as Fig.~\ref{fig:matched-SD-tau} but now for soft drop applied with angular exponent $\beta=1$.}} 
	\label{fig:matched-SD-tau-b1}
\end{figure}

\new{It is also interesting to study the behaviour of the soft-drop distribution for different values of the angular exponent $\beta$. This is done in Fig.~\ref{fig:matched-SD-tau-b1} for $\beta=1$. Because any $\beta>0$ leaves a residual double-logarithmic behaviour at small $\tau$, we see that the resummation has a bigger effect compared to the $\beta=0$ case, even for $\zc=0.1$. Furthermore, as discussed in section~\ref{sec:MC}, these distributions have larger non-perturbative corrections in the fitting region. On the other hand, we note that for $\beta=1$ the LO discontinuity is smaller and, consequently, the impact of the dilogarithm correction is reduced with respect to the $\beta=0$ case.}

Finally, it can be seen that all soft-drop distributions showed here still suffer from a discontinuity at the transition point, despite the treatment previously discussed.
This undesired feature has been pushed one order higher in perturbation theory, i.e.\ $\mathcal{O}\(\as^2\)$ but it is still sizeable. 
It originates from a discontinuity at the transition point in the second order derivative contribution, which appears because of the treatment of multiple-emission contributions.
In Ref.~\cite{Larkoski:2014wba}, this effect was smoothed out by considering finite differences rather than derivatives. Here, we prefer to leave this discontinuity apparent to stress the fact that a more rigorous solution is needed in order to use groomed distributions for precision predictions. We see this as the main challenge to be addressed in the near future.


\subsection{Towards precision}\label{sec:precision}
\begin{figure}
	\centering
	\includegraphics[width=0.45\textwidth]{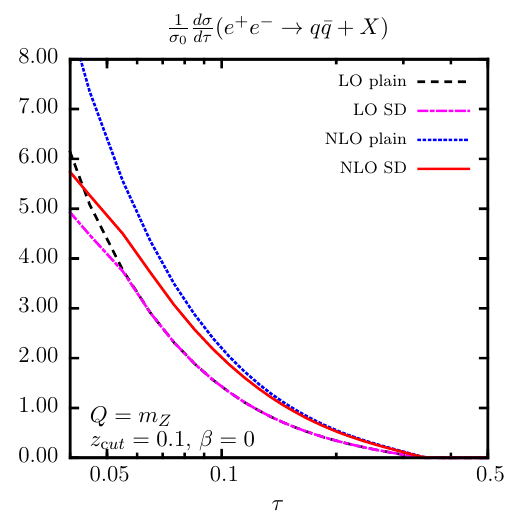}
	\includegraphics[width=0.45\textwidth]{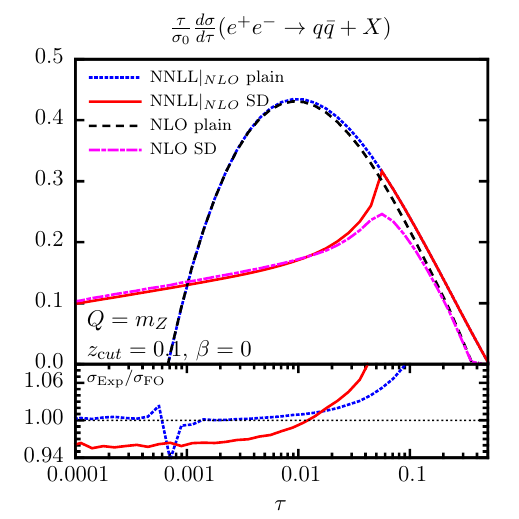}
	\caption{On the left, the thrust distribution, with and without soft drop, in fixed-order perturbation theory, namely at LO and NLO. On the right, comparisons of fixed-order results to the expansion of the NNLL resummation.} 
	\label{fig:precision}
\end{figure}
\begin{figure}
	\centering
	\includegraphics[width=0.45\textwidth]{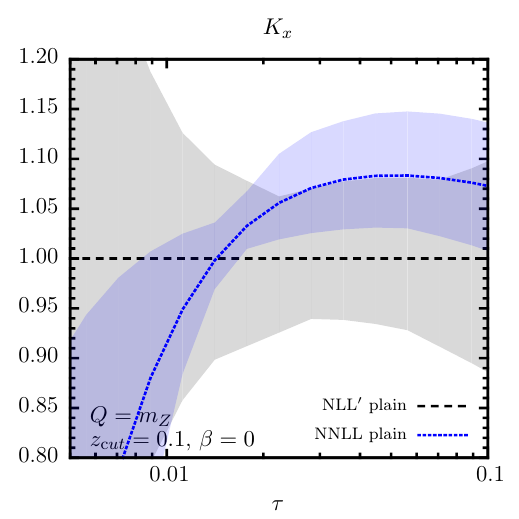}
	\includegraphics[width=0.45\textwidth]{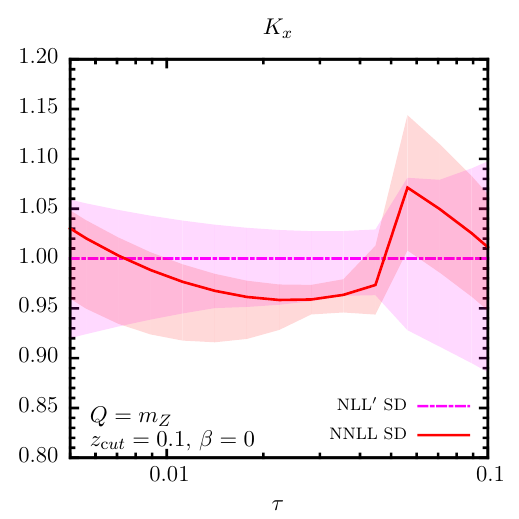}
	\caption{Ratios of the all-order distribution computed at NNLL to its NLL$'$ counterpart, on the left for plain thrust and, on the right, for soft-drop thrust.} 
	\label{fig:prec-k-factors}
\end{figure}

The results of the previous section, although interesting on their own, were not presented at a high-enough accuracy to have the ability of leading to a competitive extraction of the strong coupling. In this section, we discuss the ingredients of the calculation that need to be improved in order to reach the target accuracy. 
Let us start with the discussion of the fixed-order contribution, which we have thus far considered only at tree-level. Using the code \texttt{EVENT2}~\cite{Catani:1996jh,Catani:1996vz} we are able to obtain NLO predictions for the thrust distribution, with and without soft drop. We show the results in Fig.~\ref{fig:precision}, on the left, where we plot LO and NLO distributions on a linear scale, in order to emphasise the medium-to-large $\tau$ region. We have chosen to show the representative case $\beta=0$, $\zc=0.1$.
We can see that the NLO corrections are indeed sizeable. However, we note that they are well-reproduced by matching to the resummed calculation.
The other interesting feature that this plot shows is the effect of fixed-order corrections on the transition region. While the transition is rather sharp at LO, it becomes broader at NLO because phase-space conditions are less constraining in the presence of two emissions. Furthermore, although not so visible on the plot, the end-point of the distribution does change in going from LO to NLO.
It should be noted that $e^+e^-$ event shapes have been computed to NNLO accuracy~\cite{GehrmannDeRidder:2007jk,Gehrmann-DeRidder:2007nzq,GehrmannDeRidder:2007hr,Weinzierl:2009ms,Weinzierl:2010cw,DelDuca:2016ily}. However, the implementation of soft drop in those numerical codes is not straightforward and it is currently a work in progress. 

Second, we move to the all-order part of the calculation and we consider the effect of NNLL resummation. In Fig.~\ref{fig:precision}, on the right, we compare the expansion fo the NNLL resummation to the fixed-order results. At asymptotically small values of $\tau$, we find that the expansion of the resummation reproduces the NLO, in both cases (as in the case of Fig.~\ref{fig:exp-SD-tau} the soft-drop curve lacks finite-$\zc$ corrections).
We note that the irregular behaviour just before 10$^{-3}$ is a numerical artefact due to the fact that the denominator crosses zero in that region.
At intermediate values of $\tau$, i.e.\ in the transition region, we see that the expansion of the soft-drop resummation does not agree very well with its fixed-order counterpart. 
This is because the calculation in this region does not reach the required accuracy as it is based on the lower-order analysis which led to Eq.~(\ref{trans-rc}).
This confirms once again that in order to achieve reliable predictions across the entire range of $\tau$, accuracy in both resummation and fixed-order is not enough because soft-drop requires a detailed understanding of the region $\tau \sim \zc$.
For this reason, we prefer not to show matched NNLL+NLO results because they can be misleading until a deeper understanding of the transition region is reached. It is nevertheless interesting to show the impact of NNLL resummation, which provides the dominant effect for $\tau \ll \zc$. We study this in Fig.~\ref{fig:prec-k-factors}, where we show the ratio of the NNLL result to its NLL$'$ counterpart, on the left for plain thrust and on the right for soft-drop thrust. 
We see that the numerical impact of higher-order resummation is much reduced in the case of soft-drop thrust. Furthermore, the theoretical uncertainty, as measured by scale variation, is also significantly smaller.

\section{Resummation for jet-mass observables}\label{sec:Jet-mass}

We have seen that contributions that are not easy to control play an important role in the transition region for soft-drop thrust. 
This situation is not ideal because it makes it harder to achieve high precision in a region which is extremely relevant for phenomenology.
It is therefore interesting to analyse different observables, which share with the thrust distribution the behaviour in the soft and collinear region but that might exhibit better properties in the transition region.
The observable we consider is the jet mass. First, we will treat it similar to thrust by making use of the jet mass of a hemisphere as an observable, which is the variable that was also discussed in~\cite{Frye:2016aiz}. Second we consider the invariant mass of an anti-$k_t$~\cite{Cacciari:2008gp} jet with radius $R$ in order to assess the effect of a jet clustering radius on the transition region. 
Jet masses have been already considered in previous soft drop studies~\cite{Larkoski:2014wba,Marzani:2017mva,Marzani:2017kqd}.

\subsection{Hemisphere jet invariant mass} \label{sec:hem}
We start by considering the hemisphere mass. 
In this case, we cluster an event into exactly two jets and we look at the largest value of:
\begin{equation}
e_2^{(2)}=\frac{m_J^2}{E_J^2},
\end{equation}
with $m_J$ the jet mass and $E_J$ its energy. This is the same observable that was considered in Ref.~\cite{Frye:2016aiz}. 
Therefore, the results can be largely reused, with a slight modification due to the different definition of soft drop, which corresponds to $\zc\to\zc 2^{-\beta/2}$. 
Factorisation of the distribution in terms of hard, soft and jet functions leads to the identification of the following scales
\begin{eqnarray}
\mu_H &=& Q, \quad 
\mu_J^2 = \frac{Q^2}{4 \bar{N}}, \nonumber \\
\mu_{S_G} &=& 2^{\beta/2} Q \zc, \quad
\mu_{S_C}=\[\frac{\zc}{2^{\beta/2}\bar{N}^{\beta+1}}\]^{\frac{1}{\beta+2}} \frac{Q}{2}.
\end{eqnarray}
Note that these scales only differ in factors of two compared to the computation for thrust, since these observables share soft and collinear behaviours. Furthermore, this leads to the same anomalous dimensions.
Just as for the scales, the transition point contribution is also the same as for thrust after the change $\bar N \to 4 \bar N$ or in $\tau$-space $\tau \to e_2^{(2)}/4$. This leads to a transition contribution:
\begin{equation} \label{hem_mass:trans}
\mathcal{T}\(\tau,\zc\)=\frac{\as}{\pi}C_F\(\beta+2\) {\rm Li}_2\[\frac{1}{2}\(\frac{e_2^{(2)}}{2\;\zc} \)^{\frac{2}{\beta+2}}\].
\end{equation}

Because the resummation of the hemisphere mass was discussed in great detail in Ref.~\cite{Frye:2016aiz}, in this section we limit ourselves to an analysis of its behaviour in the transition region in order to understand whether it suffers from the same issues as the thrust. 
In Ref.~\cite{Frye:2016aiz} the computation in the small $e_2^{(2)}$ region was extended beyond the transition point. Additive matching with the fixed-order calculation was used, relying on the assumption that the resummation and its expansion cancel one another near the transition region. 
Here, we compare that procedure to ours, namely we considered the resummation of the groomed and ungroomed hemisphere mass merged together with the transition contribution Eq.~(\ref{hem_mass:trans}) and subsequently matched to fixed-order. 

\begin{figure}
	\centering
	\includegraphics[width=0.45\textwidth]{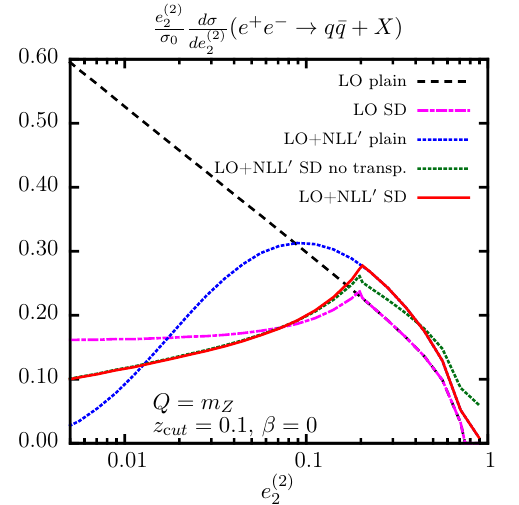}
	\includegraphics[width=0.45\textwidth]{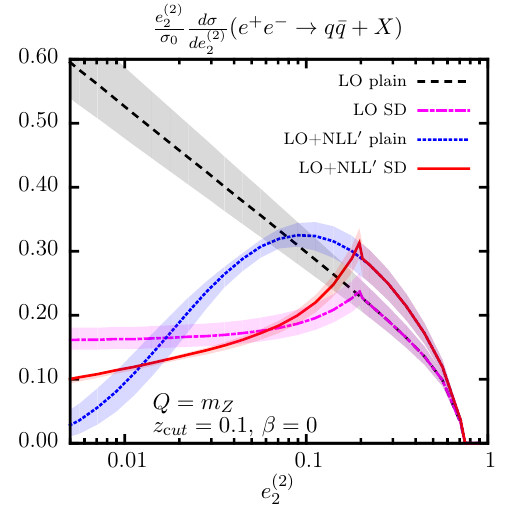}
	\caption{The results for the fixed order and resummed differential cross section of the plain and soft-dropped $e_2^{(2)}$ for $e^+ e^-$ collisions at a centre of mass energy $Q=m_Z$ and the soft drop parameters $\zc=0.1$ and $\beta=0$. The figure shows LO for plain $e_2^{(2)}$ (black dashed) and soft-dropped $e_2^{(2)}$ (magenta dashed-dotted) and the matched LO+NLL$^\prime$ cross section for plain $e_2^{(2)}$ (blue dotted) and soft-dropped $e_2^{(2)}$ (red solid) In addition the resummation without taking into account transition point effects is shown (green dotted).  Right shows the end-point corrections included with the bands from resummation and scale uncertainties.} 
	\label{fig:matched-SD-e2-no-end-point}
\end{figure}

This comparison can be seen in Fig.~\ref{fig:matched-SD-e2-no-end-point}, on the left. 
The result that makes use of the technique described in~\cite{Frye:2016aiz} is shown in the dotted green here. 
In order to make the comparison more explicit, we will not make use of the end-point modification of the logarithms previously discussed.
When compared to the solid red, which shows the result including the transition point effects, it can be seen that they agree quite well across the whole spectrum. However it can also be seen, when comparing it to the fixed order result (magenta dashed-dotted), that at this accuracy the resummation and expansion do not cancel near the transition point. Instead what is happening is that the discontinuity at the transition point is significantly canceled between the resummation and expansion. This cancellation did not happen in the case of the thrust. The difference between these observables is that for $e_2^{(2)}$ at the transition point resummation effects are significantly smaller than for thrust. This can be understood due to the fact that for $e_2^{(2)}$ the transition point is at a factor 4 larger value, which makes the logarithms significantly less important.

Finally, in Fig.~\ref{fig:matched-SD-e2-no-end-point}, on the right, we present our final results for the resummed and matched distributions, in comparison with the fixed-order ones. This is done for both the plain and the groomed hemisphere mass. For this plot, we have adopted our end-point prescriptions and we have also included uncertainty bands, which have been computed varying renormalisation and resummation scales, as previously described. Higher-order transition effects are still present but they are much reduced compared to the soft-drop thrust distributions in Fig.~\ref{fig:matched-SD-tau} and they are now within the theoretical uncertainty. 

\subsection{Narrow jet invariant mass}
We have previously performed the calculation assuming two hemisphere jets, however it is also possible to make use of a clustering algorithm without fixing the number of jets. In this case we will be making use of anti-$k_t$ clustering with a jet radius $R$. Here we define the normalisation of the observable as
\begin{equation}
\rho=\frac{m_J^2}{2 E_J^2 \(1-\cos R\)}
\end{equation}
The same computations as those performed in Section~\ref{Sec:Analytics} can be repeated for this observable. Other than the observable itself, a couple of alterations need to be made to the method of computing these different contributions. First the jet radius needs to be included in the soft drop condition. The more significant change in the one-gluon calculation is the additional condition $\theta<R$, with $\theta$ being the angle between the emission and the particle it is emitted from. For collinear emissions this condition is always satisfied. 

We start by considering the $\mathcal{O}(\as)$ contribution in the small-$R$ limit but keeping the full $\rho$ dependence
\begin{align}
\frac{1}{\sigma_0}\frac{d\sigma}{d\rho}&=\delta\(\rho\)
+C_F\frac{\as}{2\pi}\Bigg\{-4\(\frac{\log \rho}{\rho}\)_+ -\left [3\sqrt{1-4\rho}-8\log\(1+\sqrt{1-4\rho}\)+8\log 2\right]\(\frac{1}{\rho}\)_+
\nonumber\\&+\delta\(\rho\)\left(-\frac{9}{2}+\frac{2\pi^2}{3}\right)\Bigg\},
\end{align}
which, in the small $\rho$ limit, results in
\begin{equation}
\frac{1}{\sigma_0}\frac{d\sigma}{d\rho}=\delta\(\rho\)+C_F\frac{\as}{2\pi}\left[-4\(\frac{\log \rho}{\rho}\)_+ -3\(\frac{1}{\rho}\)_++\delta\(\rho\)\left(-\frac{9}{2}+\frac{2\pi^2}{3}\right)\right].
\end{equation}
When writing down the factorisation theorem, we have to pay attention to the way we treat the out-of-jet region.
However, we have just computed the full $\delta\(\rho\)$ term we can thus predict the out of jet contribution by subtracting the hard, collinear and soft expressions. In addition we know the IR divergences should cancel. Given the hard, soft and collinear functions 
\begin{eqnarray}
H&=&1+C_F\frac{\as}{2\pi} \(\frac{\mu^2}{Q^2}\)^{\epsilon}\[-\frac{2}{\epsilon^2}-\frac{3}{\epsilon}+\frac{7\pi^2}{6}-8\],\\
\tilde{J}\(\frac{\mu}{\mu_J}\)&=&C_F\frac{\as}{2\pi}\[1+\frac{\pi^2}{12}\frac{\partial^2}{\(\partial\log N\)^2}\]\(\frac{4\bar N\mu^2}{Q^2R^2}\)^{\epsilon}\[\frac{2}{\epsilon^2}+\frac{3}{2\epsilon}+\frac{1}{2}\(7-\pi^2 \) \],\\
\tilde{S}\(\frac{\mu}{\mu_S}\)&=&C_F\frac{\as}{2\pi}\[1+\frac{\pi^2}{12}\frac{\partial^2}{\(\partial\log N\)^2}\]\(\frac{4\bar N^2\mu^2}{Q^2R^2}\)^{\epsilon}\[-\frac{2}{\epsilon^2}+\frac{\pi^2}{6} \],
\end{eqnarray}
we can predict the out of jet contribution~\footnote{\new{Alternatively, the out-of-jet contribution could be computed directly using, for instance, the formalism developed in Refs.~\cite{Ellis:2010rwa,Chien:2015cka}.}}:
\begin{equation}
O=C_F \frac{\as}{2\pi}\(\frac{2\mu^2}{Q^2R}\)^\epsilon \[\frac{4\log\(\frac{R}{2}\)}{\epsilon}+\frac{\pi^2}{3}-\frac{7}{2}+6\log\(\frac{R}{2}\)\].
\end{equation}
Now this can be combined with the hard function, since both are made up of exclusively $\delta\(\rho\)$ terms:
\begin{equation}
O+H=1+C_F\frac{\as}{2\pi} \(\frac{4\mu^2}{Q^2R^2}\)^{\epsilon}\[-\frac{2}{\epsilon^2}-\frac{3}{\epsilon}+\frac{3\pi^2}{2}-\frac{23}{2}\].
\end{equation}
Whether or not an emission falls outside of the jet is independent of soft drop and only depends on the clustering algorithm. Therefore the out-of-jet function can also be applied to soft drop resummation without having to recompute it. This results in the functions for soft drop
\begin{eqnarray}
S_G\(\frac{\mu}{\mu_{S_G}}\)&=&C_F\frac{\as}{2\pi}\(\frac{2\,\mu}{\zc  \,Q\,R}\)^{2\epsilon}\frac{1}{\beta+1}\[\frac{2}{\epsilon^2}-\frac{\pi^2}{6}\],\\
\\
\tilde{S}_C\(\frac{\mu}{\mu_{S_C}}\)&=&C_F\frac{\as}{2\pi}\[1+\frac{\pi^2}{12}\frac{\partial^2}{\(\partial\log N\)^2}\]\(\frac{2\,\mu \bar{N}^{\frac{\beta+1}{\beta+2}}}{\zc^{\frac{1}{\beta+2}}  Q R}\)^{2\epsilon}\frac{\beta+2}{\beta+1}\[-\frac{1}{\epsilon^2}+\frac{\pi^2}{12}\],
\end{eqnarray}
where the hard, which includes out of jet emissions now, and the collinear functions are the same as without soft drop. The scales are given by
\begin{eqnarray}
\mu_H &=& \frac{QR}{2}, \quad
\mu_J^2 = \frac{1}{\bar{N}}\(\frac{QR}{2}\)^2,\nonumber \\
\mu_{S_G} &=& \zc \frac{QR}{2}, \quad
\mu_{S_G}=\[\frac{\zc}{\bar{N}^{\beta+1}}\]^{\frac{1}{\beta+2}} \frac{QR}{2},
\end{eqnarray}
and the anomalous dimensions remain the same as for thrust. This shows that the natural central scale for this observable is $\mu=\frac{QR}{2}$.

A crucial difference when we introduce a (small) jet-radius is the transition point contribution. What can be seen is that at the transition point, $\rho = \zc$ or $\bar{N}=1/\zc$ in Laplace space, the sum of the collinear soft and wide-angle soft functions result exactly in the soft function for the ungroomed distribution. The reason for this can be found by looking at the similar expressions without taking the small-$R$ or small-$\rho$ limit:
\begin{equation}\label{eq:finite-R-diff}
\(2+\beta\)\Li_2\[\frac{1-\cos R}{2}\(\frac{\rho} \zc\)^{\frac{2}{2+\beta}}\]-\(2+\beta\)\Li_2\[\frac{1-\cos R}{2}\]
\end{equation}
At the transition point these two terms cancel. If we approximate this in the small $\rho$ limit the first term vanishes, while the other does not. Thus, if the jet radius $R$ is large, one encounters the same transition-point issues previously discussed. Indeed the situation is analogous to the hemisphere-jet case, which can be recovered by setting $\cos R= 0$.
On the other hand, if we take the small-$R$ limit of Eq.~(\ref{eq:finite-R-diff}), then both contributions vanish up to power corrections in the jet radius. 

\begin{figure}
	\centering
	\includegraphics[width=0.45\textwidth]{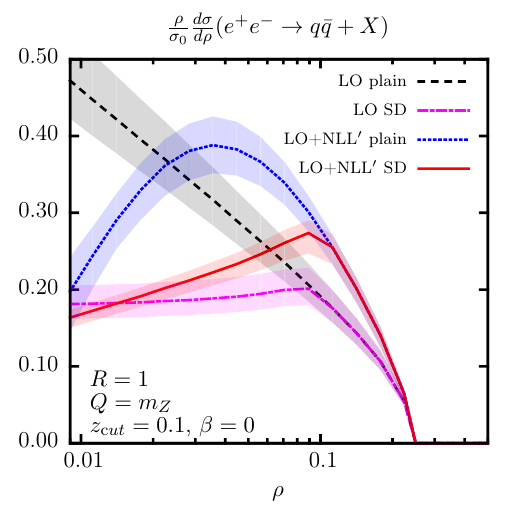}
	\includegraphics[width=0.45\textwidth]{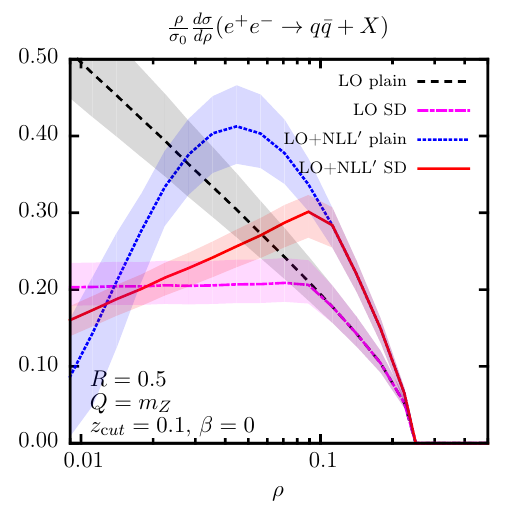}
	\caption{The results for the fixed order and resummed differential cross section of the plain and soft-dropped jet mass with $R=1$ (left) and $R=0.5$ (right) for $e^+ e^-$ collisions at a centre of mass energy $Q=m_Z$ and the soft drop parameters $\zc=0.1$ and $\beta=0$. The figure shows LO for plain $\rho$ (black dashed) and soft-dropped $\rho$ (magenta dashed-dotted) and the matched LO+NLL$^\prime$ cross section for plain $\rho$ (blue dotted) and soft-dropped $\rho$ (red solid) with the bands from resummation and scale uncertainties included.} 
	\label{fig:matched-SD-rho}
\end{figure}

Now that it has been established that small-$R$ jets are a means of suppressing the dilogarithmic transition effect, we can show the resummation for this observable. Here we make use of the LO end-point value $\rho_{\rm max}=1/4$ and we consider to jet radii, namely $R=0.5$ and $R=1.0$.
The results are shown in Fig.~\ref{fig:matched-SD-rho}.
It can be seen that for the jet mass transition point effects are small. The fixed order shows in almost constant behaviour in the region where soft drop is active and the matched cross section lines up well. This shows that it is possible to make use of a jet radius instead of making use of hemispheres in order to reduce the effects of the transition region.

\section{Conclusions}\label{sec:conclusion}
Event-shapes in $e^+e^-$ collisions are a powerful way to inspect QCD radiation in a relatively clean environment. For this reason, they have often been employed in precision QCD studies and, in particular, in determinations of the strong coupling constant.
Despite the fact that IRC safety guarantees that non-perturbative corrections due to the hadronisation process are power-suppressed, these have a non-negligible impact on event-shape distributions in region of phase-space where many data points live. In particular, fits to determine the strong coupling show a significant correlation between $\as$ and non-perturbative parameters. 

In this paper, we have put forward the idea of using techniques developed in the context of jet substructure to reduce an observable sensitivity to non-perturbative physics. 
In particular, we have considered the soft-drop algorithm and we have applied it to the event-shape thrust.
We have first performed a study using Monte Carlo parton shower simulations and we have found that the impact of these non-perturbative corrections is significantly reduced when soft drop is applied.
This opens up the possibility of performing fits for the strong coupling that rely on a wider region of phase-space where hadronisation corrections are genuinely small. 
In this context, we have shown that the effectiveness in reducing non-perturbative corrections, at the energy considered here $Q=m_Z$, quickly degrades as the angular exponent $\beta$ increases, and preferred options appear to be $\beta=0$ and, perhaps, $\beta=1$. On the other hand, we have found that the dependence on the energy threshold $\zc$ is less pronounced. Thus, a mild energy cutoff, e.g.\ $\zc=0.05$, appears to be an promising compromise between reducing hadronisation corrections, while maintaining the bulk of the dataset.

In order for this enterprise to be successful, reduction in non-perturbative effects must be accompanied by our ability of performing perturbative calculations for soft-drop event shapes with an accuracy that matches the one for traditional event shapes. 
While this is certainly possibile for fixed-order calculations, work has to be done in the context of resummation where soft-drop observables are currently known to NNLL, while the un-grommed thrust distribution is known one order higher, namely N$^3$LL.
Furthermore, as we have pointed out in this study, complications may arise in the description of the so-called transition region, i.e.\ the region of phase-space where soft drop starts to become active.
The hierarchy of scales that characterises the deep infra-red and collinear region does not apply here and one becomes sensitive to a new class of contributions. These were investigated here to first order, but a more detailed analysis is necessary if we want to maintain perturbative accuracy in this region.
However, our analysis also shows that observables that share the same behaviour in the soft/collinear limit, may have exhibit very different sensitivity to these contributions.
From this point of view, we have found soft-drop thrust to be particularly sensitive to these corrections, which are instead parametrically suppressed if we choose to measure the invariant mass of jet with radius $R$.

In conclusion, grooming algorithms such as soft drop show a promising reduction in the non-perturbative corrections, even when applied to $e^+e^-$ collision at the $Z$ pole, with potential benefits for $\as$ determination, provided that the perturbative structure of the resulting distribution is under theoretical control in the range relevant for phenomenology.
Furthermore, we note that the recursive structure of the soft-drop algorithm opens up new ways of defining event-shape or observables.
The traditional jet mass, and thrust, receives important (i.e.\ NLL) contributions from any number of un-ordered emissions in the final state, while it is possible to define observables on the two prongs that first pass the soft-drop condition. Two-pronged observables exhibit different sensitivity to ``multiple-emissions" that might lead to a simplification of their all-order treatment, while directly exposing the strong coupling at the tagged splitting. 
This study is part of a rather ambitious project which aims to apply techniques developed for searches to precision measurements (see Ref.~\cite{Hoang:2017kmk} for work in the context of top-quark mass extraction) and we look forward to continuing working in this direction.

\begin{acknowledgments}
We thank Andrew Larkoski, Gregory Soyez and Jesse Thaler for many useful discussions and Steffen Schumann for his comments on this manuscript.
This work was partly supported by the the U.S.\ National Science Foundation, under grant PHY-1619867, All-Order Precision for LHC Phenomenology. 
The work leading to this publication was also supported by the German Academic Exchange Service (DAAD) with funds from the German Federal Ministry of Education and Research (BMBF) and the People Programme (Marie Curie Actions) of the European Union Seventh Framework Programme (FP7/2007-2013) under REA grant agreement n.\ 605728 (P.R.I.M.E.\ Postdoctoral Researchers International Mobility Experience).
\end{acknowledgments}

\appendix

\section{Transition points in the soft-drop thrust distribution}\label{sec:2nd-trans}
In this appendix we consider three-parton configuration and we study, to lowest order in the strong coupling, the kinematic configurations that give rise to transition points in the thrust distribution. We focus on the $\beta=0$ case and
we start by considering a configuration with three massless partons:
\begin{align}
E_1+E_2+E_3&=Q, \nonumber\\
\vec{p}_1+\vec{p}_2+\vec{p}_3&=0, \nonumber\\
E_i=|\vec{p}_i|, \quad i&=1,2,3.
\end{align}
 The thrust axis \emph{before} soft-drop divides the event into two hemispheres and without loss of generality, we assume that partons 2 and 3 are in the same hemisphere, while parton 1 is recoiling against them in the opposite one. Thus we have $ \vec{p}_2 \cdot \vec{p}_3 \ge 0$ and
 \begin{equation}
 E_1^2=E_2^2+E_3^2+ 2 \vec{p}_2 \cdot \vec{p}_3\ge E_2^2+E_3^2,
 \end{equation}
 which implies $E_1\ge E_2, E_3$. 
 We then apply soft drop and we consider the situation in which $E_3<E_2$ and parton 3 is just groomed away or just passes. For the $\beta=0$ case this happens if 
 \begin{equation}
\frac{E_{3}}{E_2+E_3} =z_{c} \Longrightarrow E_{3}=\frac{z_{c}}{1-z_{c}}E_{2}.
\end{equation}
\subsection{The asymptotic region of small $\tausd$}
In order to determine the resulting $\tausd$ when parton 3 is groomed away, the algorithm then calculates thrust on the two-parton event. However, the partons are not back-to-back and this results in a non-trivial configuration for the soft-drop thrust axis and, consequently, a non-zero value of $\tausd$. We have
\begin{align}
\tausd&= 1- \max_{\vec{n}}\frac{|\vec{p}_1 \cdot \vec{n}|+|\vec{p}_2 \cdot \vec{n}|}{|\vec{p}_1|+|\vec{p}_2|}.
\end{align}
The thrust axis is found to be
\begin{equation}
\vec{n}_\textsc{SD}=\frac{\vec{p}_1-\vec{p}_2}{|\vec{p}_1-\vec{p}_2|},
\end{equation}
which leads to
\begin{equation}
\tausd=1-\frac{\sqrt{\left(\vec{p}_{1}-\vec{p}_{2}\right)^{2}}}{E_{1}+E_{2}}=\frac{E_{1}+E_{2}-\sqrt{E_{1}^{2}+E_{2}^{2}-2\vec{p}_{1}\cdot\vec{p}_{2}}}{E_{1}+E_{2}}=\frac{E_{1}+E_{2}-\sqrt{2E_{1}^{2}+2E_{2}^{2}-E_{3}^{2}}}{E_{1}+E_{2}},
\label{eq:thrust2}
\end{equation}
with
\begin{equation}
E_{3}^{2}=\left|\vec{p}_{3}\right|^{2}=E_{1}^{2}+E_{2}^{2}+2\vec{p}_{1}\cdot\vec{p}_{2}
\end{equation}
Note that Eq.~(\ref{eq:thrust2}) vanishes in the soft limit $E_3 \to 0$ because $E_1 \to E_2$. Therefore, at least to this order in perturbation theory, $\tausd$ is infra-red and collinear safe.

We are interested in finding the maximum value of $\tausd$ which is sensitive to this type of kinematic configurations. In the main text, we argued that this should be $
\tausd=\mathcal{O}\left( z_c^2\right)$. We can now make a more quantitative statement. The maximum of Eq.~(\ref{eq:thrust2}) is reached when $E_3$ is as large as possible, while being groomed away and $E_1$ and $E_2$ equally share the remaining energy $Q-E_3$. We find
\begin{align}
\bar{E}_1&= \bar{E}_2=\frac{1-z_c}{2-z_c}Q, \nonumber \\
\bar{E}_3&= \frac{z_c}{2-z_c}Q.
\end{align}
Filling these energies into Eq.~(\ref{eq:thrust2}) leads to
\begin{equation}
\bar{\tau}_\textsc{SD}=\frac{2\left(1-z_{c}\right)-\sqrt{4-8z_{c}+3z_{c}^{2}}}{2\left(1-z_{c}\right)}=
\frac{z_{c}^{2}}{8}+\mathcal{O}\left(z_{c}^{3}\right).
\end{equation}
For $z_{c}=0.1$, we find $\bar{\tau}_\textsc{SD}\approx0.00154$, which agrees with what is seen in Fig.~\ref{Fig:Fixed-order}.

In the soft emission limit $\tausd$ given by Eq.~(\ref{eq:thrust2}) is equal to
\begin{equation}
\tausd = \frac{k^{+}k^{-}}{2Q^2}.
\end{equation}
When this is effect is included it adds an additional contribution to the soft function:
\begin{align}\label{eq:soft-tausd}
\tilde{S}\(N, \zc, \beta\)&=g_s^2C_F\(\frac{\mu^2e^{\gammae}}{4\pi}\)^{\epsilon}\int d\tausd\; e^{-N\tau}\int\frac{d^{4-2\epsilon}k}{\(2\pi\)^{3-2\epsilon}}\frac{n\cdot \bar{n}}{k^- k^+}\delta\(k^2\)\Theta\(k^-+k^+\)\Theta\(k^--k^+\)\nonumber\\
&\times\Theta\(\zc Q\[\frac{k^+}{k^0}\]^{\beta/2}-2k^0\)\[\delta\(\tausd-\frac{k^-k^+}{2Q^2}\)-\delta\(\tausd\)\]+\(n\leftrightarrow\bar{n}\)\nonumber\\
&=C_F\frac{\as}{2\pi}\[-\log^2\tausd+2\log\tausd\log\(\frac{\zc^2}{2}\)+\log^2\(\frac{\zc^2}{2}\)-\frac{\pi^2}{3}\]+\Ord\(\tausd\),
\end{align}
if $\tausd<\zc^2/8$. Hence, we find a double logarithmic enhancement for the differential cross section in the small $\tausd$ limit, even for $\beta=0$. For this reason, we have decided to discard the naive version of soft-drop thrust $\tausd$ in favor of the better behaved $\tausd'$.

\subsection{Transition point at large $\tausd$}
The same kinematic configuration can be applied to the case where no particle is groomed away in order to obtain the transition point above which the distribution returns to the un-groomed thrust differential cross section.
Since no particles are groomed away conservation of momentum can now be applied:
\begin{align}
\tausd&= 1- \max_{\vec{n}}\frac{|\vec{p}_1 \cdot \vec{n}|+|\vec{p}_2 \cdot \vec{n}|+|\vec{p}_3 \cdot \vec{n}|}{|\vec{p}_1|+|\vec{p}_2|+|\vec{p}_3|}.
\end{align}
The thrust axis is found to be
\begin{equation}
\vec{n}=\frac{\vec{p}_1-\vec{p}_2-\vec{p}_3}{|\vec{p}_1-\vec{p}_2-\vec{p}_3|}=\frac{\vec{p}_1}{|\vec{p}_1|},
\end{equation}
which leads to
\begin{equation}
\tausd=1-\frac{\sqrt{\left(\vec{p}_{1}-\vec{p}_{2}-\vec{p}_3\right)^{2}}}{E_{1}+E_{2}+E_{3}}=\frac{Q-2E_{1}}{Q}.
\label{eq:thrust3}
\end{equation}
Filling the energy derived in the previous section into this equation leads to
\begin{equation}
\bar{\tau}=\frac{z_{c}}{2-z_{c}}=
\frac{z_{c}}{2}+\mathcal{O}\left(z_{c}^{2}\right).
\end{equation}
For $z_{c}=0.1$, this leads to $\bar{\tau}\approx0.05263$, which also agrees with what is seen in Fig.~\ref{Fig:Fixed-order}.

\section{Details of the analytic calculation}\label{app:details}
\subsection{Scales and coefficients}
In order to compute the scales for each of the factorized functions and their associated anomalous dimensions we will follow the derivation of Appendices B-E of~\cite{Frye:2016aiz}. For the computations we will be making use of light-cone coordinates defined by $n^\mu$ the jet direction and $\bar n^\mu$ the opposite direction resulting in $k^-=\bar{n}\cdot k$, $k^+=n\cdot k$ and $k_\bot$ the components transverse to $n$.

\subsubsection{Collinear function}
The collinear (or jet) function can be derived using standard splitting functions, which describes the emission of a particle in the collinear limit. This calculation is observable dependent, however it is not groomer dependent. Therefore the same results as for un-groomed thrust can be used here~\cite{Bosch:2004th}. For the collinear limit we can assume $k^-\gg k^+$, which implies $k^{(0)}\approx k^-/2$.

For the following we will compute the expressions in Laplace space with the Laplace space conjugate $N$:
\begin{equation}
\tilde{J}\(\nu\)=\int_{0}^{\infty}d\tau e^{-N\tau} J\(\tau\)
\end{equation}
In Laplace space the one-loop Jet function in the $\MSbar$ scheme is given by:
\begin{eqnarray}
\tilde{J}\(\frac{\mu}{\mu_J}\)&=&g_s^2\(\frac{\mu^2e^{\gammae}}{4\pi}\)^{\epsilon}\int d\tau\; e^{-N\tau}\int\frac{d^{4-2\epsilon}k}{\(2\pi\)^{3-2\epsilon}} \frac{P_{qg}\(z\)}{k^+ Q}  \;\delta\(\tau-\frac{k^+}{z Q} \)\delta\(k^2\)\Theta\(k^-\)\nonumber\\
&=&g_s^2\frac{\Omega_{1-2\epsilon}}{4\(2\pi\)^{3-2\epsilon}}\(\frac{\mu^2e^{\gammae}}{Q4\pi}\)^{\epsilon}\int d\tau\; e^{-N\tau} \int \frac{dz}{\(1-z\)^\epsilon} \int dk^+ \(k^+\)^{-1-\epsilon} P_{qg}\(z\)\;\delta\(\tau-\frac{k^+}{zQ} \)\nonumber\\
&=&C_F\frac{\as}{2\pi}\[1+\frac{\pi^2}{12}\frac{\partial^2}{\(\partial\log N\)^2}\]\(\frac{\bar N\mu^2}{Q^2}\)^{\epsilon}\[\frac{2}{\epsilon^2}+\frac{3}{2\epsilon}+\frac{1}{2}\(7-\pi^2 \) \],
\label{eq:Jet-func}
\end{eqnarray}
where 
we have used the integration variable $(1-z)=k^-/Q$ and introduced $\bar N = N e^{\gammae}$ which results from the Mellin transform approximation from Appendix~A of~\cite{Catani:2003zt} that also applies to Laplace transformations up to $\Ord(1/N)$. 
This derivation is for one of the hemisphere jets, the other jet can be computed in a similar manner.

\subsubsection{Global soft function}
Since soft wide angle emissions fail the soft drop condition in general this function will not depend on the observable and will only depend on the grooming condition. Therefore this result will be the same as the one presented in~\cite{Frye:2016aiz}.
\begin{eqnarray}
S_G\(\frac{\mu}{\mu_{S_G}}\)&=&g_s^2C_F\(\frac{\mu^2e^{\gammae}}{4\pi}\)^{\epsilon}\int\frac{d^{4-2\epsilon}k}{\(2\pi\)^{3-2\epsilon}}\frac{n\cdot \bar{n}}{k^- k^+}\delta\(k^2\)\Theta\(k_0\)\Theta\(k^--k^+\)\Theta_{\mathrm{SD}}+\(n\leftrightarrow\bar{n}\)\nonumber\\
&=&g_s^2C_F\(\frac{\mu^2e^{\gammae}}{4\pi}\)^{\epsilon}\frac{\Omega_{1-2\epsilon}}{\(2\pi\)^{3-2\epsilon}}\int \frac{dk^+}{\(k^+\)^{1+\epsilon}} \int\frac{dk^-}{\(k^-\)^{1+\epsilon}}\Theta\(k^0\)\Theta\(k^--k^+\)\Theta_{\mathrm{SD}}\nonumber\\
&=&\frac{\as C_F}{2\pi}\(\frac{\mu}{2^{\beta/2} \zc  Q}\)^{2\epsilon}\[\frac{2}{\beta+1}\frac{1}{\epsilon^2}-\frac{\pi^2}{6}\(\frac{1}{1+\beta}+2+\beta\)\].
\end{eqnarray}
with the soft drop condition defined as
\begin{equation}
\Theta_{\mathrm{SD}}=\Theta\(\zc \frac{Q}{2}\[\frac{k^+}{k^0}\]^{\beta/2}-k^0\)
\end{equation}

\subsubsection{Soft collinear function}
For the soft collinear function both the observable and the grooming method need to be taken into account. However, the anomalous dimensions and scale can be determined through the cancellation of the IR-divergences. However,  in order to find the constant contributions, we will still need to compute this function. We find
\begin{eqnarray}
\tilde{S}_C\(\frac{\mu}{\mu_{S_C}}\)&=&g_s^2C_F\(\frac{\mu^2e^{\gammae}}{4\pi}\)^{\epsilon}\int d\tau\; e^{-N\tau}\int\frac{d^{4-2\epsilon}k}{\(2\pi\)^{3-2\epsilon}}\frac{n\cdot \bar{n}}{k^- k^+}\delta\(k^2\)\Theta\(k^-\)\nonumber\\
&&\times\[\Theta\(\zc Q\[2\frac{k^+}{k^-}\]^{\beta/2}-k^-\)\delta\(\tau\)+\Theta\(k^--\zc Q\[2\frac{k^+}{k^-}\]^{\beta/2}\)\delta\(\tau-\frac{k^+}{Q}\)\]\nonumber\\
&=&-g_s^2C_F\(\frac{\mu^2e^{\gammae}}{4\pi}\)^{\epsilon}\frac{\Omega_{1-2\epsilon}}{2\(2\pi\)^{3-2\epsilon}}\int \frac{dk^+}{\(k^+\)^{1+\epsilon}} \int\frac{dk^-}{\(k^-\)^{1+\epsilon}}\Theta\(k^-\)\nonumber\\
&&\times\Theta\(k^--\zc Q\[2\frac{k^+}{k^-}\]^{\beta/2}\)\[1+\frac{\pi^2}{12}\frac{\partial^2}{\(\partial\log N\)^2}\]\Theta\(\frac{k^+}{Q} -\frac{1}{\bar N}\)\nonumber\\
&=&\frac{\as}{2\pi}C_F\[1+\frac{\pi^2}{12}\frac{\partial^2}{\(\partial\log N\)^2}\]\(\frac{\mu \bar{N}^{\frac{\beta+1}{\beta+2}}}{2^{\frac{\beta/2}{\beta+2}} \zc^{\frac{1}{\beta+2}}  Q}\)^{2\epsilon}\frac{\beta+2}{\beta+1}\[-\frac{1}{\epsilon^2}+\frac{\pi^2}{12}\].
\end{eqnarray}

\section{Resummation formulae} \label{app:resum_formulae}
As described in Section~\ref{sec:implementation}  in the soft-drop region the resummation is written in the form
\begin{align}\label{eq:Integrated-xsec}
\Sigma\(\tau\)&=\frac{1}{2\pi i}\int_C \frac{dN}{N} \[1+\sum_{n=1}^{\infty}\(\frac{\as}{\pi}\)^{n}\tilde{C}^{\(n\)},\]e^{\tilde{F}\(\lambda_{\bar{N}},\lambda_{\zc}\)},
\end{align}
where $\tilde{C}$ encapsulates the constant contributions in $\tau$ and $\zc$ an $\as$ is computed at a scale $\mu$. The resummed exponent $\tilde{F}$ is given by
\begin{equation}
\tilde{F}\(\lambda_{\bar{N}},\lambda_{\zc}\) = \frac{1}{\as}f_1\(\lambda_{\bar{N}},\lambda_{\zc}\)+f_2\(\lambda_{\bar{N}},\lambda_{\zc}\)+\as f_3\(\lambda_{\bar{N}},\lambda_{\zc}\).
\end{equation}
These functions are given by:
\begin{eqnarray}
f_1^{K}\(\lambda_T\)&=&\frac{\Gamma_K^{(0)}}{2 b_0^2\pi}\[\(1+2\lambda_T\)\log\(1+2\lambda_T\)-2\lambda_T\],\\
f_2^{K}\(\lambda_T\)&=&\frac{\Gamma_K^{(1)}}{ b_0^2\pi^2}\[\lambda_T-\frac{1}{2}\log\(1+2\lambda_T\)\]+\frac{\Gamma_K^{(0)}b_1}{ 4b_0^3\pi}\[\log\(1+2\lambda_T\)\(2+\log\(1+2\lambda_T\)\)-4\lambda_T\]\nonumber\\
&&+\frac{\Gamma_K^{(0)}}{ 2b_0\pi}\log\(1+2\lambda_T\)L_\mu^K-\frac{\gamma_K^{(0)}}{2b_0\pi}\log\(1+2\lambda_T\),\\
f_3^{K}\(\lambda_T\)&=&\frac{\Gamma_{K}^{\left(2\right)}}{b_{0}^{2}\pi^{3}}\frac{\lambda_T^{2}}{1+2\lambda_T}-\frac{\Gamma_{K}^{\left(1\right)}b_{1}}{b_{0}^{3}\pi^{2}}\left[\frac{\log\left(1+2\lambda_T\right)-2\left(1-\lambda_T\right)\lambda_T}{2\left(1+2\lambda_T\right)}\right]\nonumber \\
&&+ \frac{\Gamma_{K}^{\left(0\right)}b_{2}}{b_{0}^{3}\pi}\left[\frac{\left(1+2\lambda_T\right)\log\left(1+2\lambda_T\right)-2\lambda_T\left(1+\lambda_T\right)}{2\left(1+2\lambda_T\right)}\right]+\frac{\Gamma_{K}^{\left(0\right)}b_{1}^{2}}{b_{0}^{4}\pi}\left[\frac{\left(\log\left(1+2\lambda_T\right)-2\lambda_T\right)^{2}}{4\left(1+2\lambda_T\right)}\right]\nonumber \\
&&+ \frac{\Gamma_{K}^{\left(1\right)}}{b_{0}\pi^{2}}L^K_{\mu}\frac{\lambda_T}{1+2\lambda_T}+\frac{\Gamma_{K}^{\left(0\right)}b_{1}}{2b_{0}^{2}\pi}L^K_{\mu}\left[\frac{\log\left(1+2\lambda_T\right)-2\lambda_T}{1+2\lambda_T}\right] - \frac{\Gamma_{K}^{\left(0\right)}}{2\pi}\(L^K_{\mu}\)^{2}\frac{\lambda_T}{1+2\lambda_T}\nonumber \\
&& -\frac{\gamma_{K}^{\left(1\right)}+2\pi b_{0}K\(1\)}{b_{0}\pi^{2}}\frac{\lambda_T}{1+2\lambda_T}-\frac{\gamma_{K}^{\left(0\right)}b_{1}}{2b_{0}^{2}\pi}\left[\frac{\log\left(1+2\lambda_T\right)-2\lambda_T}{1+2\lambda_T}\right]+\frac{\gamma_{K}^{\left(0\right)}}{\pi}L^K_{\mu}\frac{\lambda_T}{1+2\lambda_T},
\end{eqnarray}
for any function $K$ with the sum given by
\begin{eqnarray}
f_i\(x,y\)&=&f_i^{S_G}\(p^{(\bar{N})}_{S_G}x+p^{(\zc)}_{S_G}y\)+2f_i^{S_C}\(p^{(\bar{N})}_{S_C}x+p^{(\zc)}_{S_C}y\)+2f_i^{J}\(p^{(\bar{N})}_{J}x+p^{(\zc)}_{J}y\)\nonumber\\
&=&f_i^{S_G}\(p^{(\zc)}_{S_G}y\)+2f_i^{S_C}\(p^{(\bar{N})}_{S_C}x+p^{(\zc)}_{S_C}y\)+2f_i^{J}\(p^{(\bar{N})}_{J}x\).
\end{eqnarray}
and
\begin{equation}
L^K_{\mu}=\log\(\frac{Q^2}{\mu^2}\)+2p^{(2)}_K\log 2
\end{equation}
We note that some of the terms cancel in the final exponential because of Eq.~(\ref{eq:power-sum}).
Finally, the coefficient $\tilde{C}^{\(n\)}$ is given by
\begin{equation}
\tilde{C}^{\(1\)}_{K}=K\(1\)+\[\log\(\frac{Q^2}{\mu^2}\)+2p^{\(2\)}_{K}\log 2\]\[-\frac{\gamma^{\(0\)}_K}{2}+\frac{\Gamma^{\(0\)}_K}{4}\(\log\(\frac{Q^2}{\mu^2}\)+2p^{\(2\)}_{K}\log 2\)\],
\end{equation}
for any function $K$ and the sum now includes the hard function
\begin{equation}
\tilde{C}^{\(n\)}=\tilde{C}^{\(n\)}_{H}+\tilde{C}^{\(n\)}_{S_G}+2\tilde{C}^{\(n\)}_{S_C}+2\tilde{C}^{\(n\)}_{J}.
\end{equation}
In principle the full result is a product of the all order contributions for the different functions, however up to order $\as$ we can write this as a sum.

\subsection{Laplace inversion}\label{app:laplace}
For the inverse Laplace transform we shall make use of a technique first proposed in~\cite{Catani:1992ua}. Furthermore, we find convenient to rewrite the inversion into a form similar to what was presented in~\cite{Monni:2011gb}.
In order to perform the Laplace inversion we expand Eq.~(\ref{eq:Integrated-xsec}) about $\bar{\nu}=\bar{N}\tau=1$:
\begin{eqnarray}
\Sigma\(\tau\)&=&\[1+\sum_{i=1}^{\infty}\(\frac{\as}{\pi}\)^{n}\tilde{C}^{\(n\)}\] e^{\tilde{F}\(-\lambda_\tau,\lambda_{\zc}\)}\frac{1}{2\pi i}\int_C \frac{d\nu}{\nu}\\
&\times&\exp\[\nu+\tilde{F}^{\(1\)}\(-\lambda_\tau,\lambda_{\zc}\)\(\log\nu+\gammae\)+\frac{1}{2}\tilde{F}^{\(2\)}\(-\lambda_\tau,\lambda_{\zc}\)\(\log\nu+\gammae\)^2+\Ord\(\as^{n+2}\log^{n}\tau\)\],\nonumber
\end{eqnarray}
where we suppress terms beyond NNLL accuracy and introduce the new integration variable $\nu=N\tau$. Here $\tilde{F}^{(n)}$ is defined as 
\begin{equation}
\tilde{F}^{(n)}\(-\lambda_\tau,\lambda_{\zc}\)=\frac{d^n }{d\log^n\frac{1}{\tau}}\tilde{F}\(\as b_0\log\frac{1}{\tau},\lambda_{\zc}\),
\end{equation}
with $b_0=\frac{11 C_A -2 n_f}{12 \pi}$. The necessary functions are, up to NNLL accuracy, given by
\begin{eqnarray}
	\tilde{F}^{(1)}\(x,y\)&=& b_0 \frac{d}{dx} f_1\(x,y\)+\as b_0 \frac{d}{dx} f_2\(x,y\)+\Ord\(\as^{n+2}\log^{n}\tau\),\nonumber\\
	\tilde{F}^{(2)}\(x,y\)&=& \as b_0^2 \frac{d^2}{dx^2} f_1\(x,y\)+\Ord\(\as^{n+2}\log^{n}\tau\).
\end{eqnarray}
This results in
\begin{eqnarray}
\Sigma\(\tau\)&=&\[1+\sum_{i=1}^{\infty}\(\frac{\as}{\pi}\)^{n}\tilde{C}^{\(n\)}\] \exp\[\tilde{F}\(-\lambda_\tau,\lambda_{\zc}\)+\gammae\tilde{F}^{\(1\)}\(-\lambda_\tau,\lambda_{\zc}\)+\frac{\gammae^2}{2}\tilde{F}^{\(2\)}\(-\lambda_\tau,\lambda_{\zc}\)\]
\nonumber \\ &\times&
\frac{1}{2\pi i}
\int_C \frac{d\nu}{\nu}\exp\Bigg[\nu+\(\tilde{F}^{\(1\)}\(-\lambda_\tau,\lambda_{\zc}\)+\gammae\tilde{F}^{\(2\)}\(-\lambda_\tau,\lambda_{\zc}\)\)\log\nu
\nonumber \\ 
&+&\frac{1}{2}\tilde{F}^{\(2\)}\(-\lambda_\tau,\lambda_{\zc}\)\log^2\nu+\Ord\(\as^{n+2}\log^{n}\tau\)\Bigg].\nonumber\\
\end{eqnarray}
This integral can be solved by making use of~\cite{Catani:1992ua}
\begin{equation}
\frac{1}{2\pi i}\int_C \frac{d\nu}{\nu}\log^n\nu\exp\[\nu+G\log\nu\]=\frac{d^k}{dG^k}\frac{1}{\Gamma\(1-G\)}.
\end{equation}
In order to change the integral into this form we are required to expand out the $\log^2\nu$ term in the exponent. The terms that are neglected here are of the order $\Ord\(\as^k\(\as\log\tau\)^n\)$ with $k\geq 2$ and $n\geq 1$. Using this method we obtain
\begin{eqnarray}
&R\(\tau\)&=\[1+\sum_{i=1}^{\infty}\(\frac{\as}{\pi}\)^{n}\tilde{C}^{\(n\)}\] \exp\[\frac{1}{\as}f_1+f_2+\gammae b_0f_1^{\prime}+\as\( f_3+\gammae b_0 f_2^{\prime}+\frac{\gammae^2}{2} b_0^2f_1^{\prime\prime}\)\]\\
\!&\times&\hspace{-1em}\frac{1}{\Gamma\(1-b_0 f_1^{\prime}\)}\[1+\as b_0 \(f_2^\prime+\gammae b_0 f_1^{\prime\prime}\)\psi^{\(0\)}\(1- b_0 f_1^{\prime}\)+\frac{1}{2}\as b_0^2 f_1^{\prime\prime}\(\psi^{\(0\)}\(1-b_0 f_1^{\prime}\)^2-\psi^{\(1\)}\(1-b_0 f_1^{\prime}\)\)\],\nonumber
\end{eqnarray}
and we have suppressed the arguments of $f_i$ and the prime indicates the derivative with respect to the first argument of $f_i$.
Finally we can exponentiate the results of the integral and combine everything in the final result:
\begin{equation}
R\(\tau\)=\[1+\sum_{i=1}^{\infty}\(\frac{\as}{\pi}\)^{n}C^{\(n\)}\] \exp\[\frac{1}{\as}g_1\(-\lambda_\tau,\lambda_{\zc}\)+g_2\(-\lambda_\tau,\lambda_{\zc}\)+\as g_3\(-\lambda_\tau,\lambda_{\zc}\)\],
\end{equation}
where
\begin{eqnarray}
g_1\(x,y\) &=& f_1\(x,y\),\nonumber\\
g_2\(x,y\) &=& f_2\(x,y\)+\gammae b_0f_1^{\prime}\(x,y\)-\log\Gamma\(1-b_0 f_1^{\prime}\(x,y\)\),\nonumber\\
g_3\(x,y\) &=& f_3\(x,y\)+b_0 f_2^\prime\(x,y\) \(\psi^{\(0\)}\(1-b_0 f_1^{\prime}\(x,y\)\)+\gammae\)+\frac{\pi^2}{12}b_0^2 f_1^{\prime\prime}\(0,0\)\nonumber\\
&&+\frac{b_0^2}{2} f_1^{\prime\prime}\(x,y\)\(\psi^{\(0\)}\(1- b_0 f_1^{\prime}\(x,y\)\)^2-\psi^{\(1\)}\(1- b_0 f_1^{\prime}\(x,y\)\)+2\gammae\psi^{\(0\)}\(1-b_0 f_1^{\prime}\(x,y\)\)+\gammae^2\)\nonumber\\
C^{(1)} &=& \tilde{C}^{(1)}-\frac{\pi^2}{12}b_0^2 f_1^{\prime\prime}\(0,0\),
\end{eqnarray}
where we have shifted the constant contribution of $g_3$ to the rest of the constant contributions in $C^{(1)}$, which means $g_3\(0,0\)=0$.

The contributions from the transition region should be taken into account somewhat differently. The dilogarithm in Laplace space was computed using an approximation that only holds for logarithms (Appendix~A of~\cite{Catani:2003zt}) and should instead be treated in $\tau$ space directly, where we neglect multiple emission contributions. Only taking into account the single emission case allows us to make use of the dilogarithm from the inclusive soft function in thrust space as presented in Eq.~(\ref{eq:soft-func-cntd}). This corresponds to the simple substitution $1/\bar{N}\to\tau$.

Since we are interested in the differential cross section the derivative of $R$ will need to be taken with respect to $\tau$:
\begin{eqnarray}
\tau\frac{d\sigma}{d\tau}\(\tau\)&=&-\[1+\sum_{i=1}^{\infty}\(\frac{\as}{\pi}\)^{n}C^{\(n\)}\] \[b_0 g_1^{\prime}\(-\lambda_\tau,\lambda_{\zc}\)+\as b_0 g_2^{\prime}\(-\lambda_\tau,\lambda_{\zc}\)+\as^2 b_0 g_3^{\prime}\(-\lambda_\tau,\lambda_{\zc}\)\]\nonumber\\
&&\times\exp\[\frac{1}{\as}g_1\(-\lambda_\tau,\lambda_{\zc}\)+g_2\(-\lambda_\tau,\lambda_{\zc}\)+\as g_3\(-\lambda_\tau,\lambda_{\zc}\)\].
\end{eqnarray}
Here at LL accuracy only $f_1$ and $f_1^\prime$ are needed, at NLL accuracy in addition we need $f_2$, $f_2^\prime$ and from the derivative of the additional terms used for the inversion $f_1^{\prime\prime}$.

In order to assess the uncertainty due to missing logarithmic orders, we rescale the argument the argument of the logarithms we are resumming by an arbitrary factor $x_L$,
\begin{equation}
\log\left(x_{L}\tau\right),\quad\log\left(x_{L}\frac{z_{\mathrm{cut}}}{2}\right),
\end{equation}
where an additional factor $1/2$ is included in the logarithm of $\zc$ in order to ensure that these logarithms are the same at the transition point.
In order to maintain NLL accuracy, we have to modify the functions $f_2$ and $\tilde{C}^{(1)}$:
\begin{equation}
p^{\left(2\right)}\log 2\to p^{\left(2\right)}\log 2+p^{\left(z_{\mathrm{cut}}\right)}\log 2-\log x_{L}\left(p^{\left(z_{\mathrm{cut}}\right)}-p^{\left(\bar{N}\right)}\right).
\end{equation}

\subsection{Treatment of the end point}\label{sec:end-point}
Since the matched cross section should be valid over the full range, it is convenient to force  the end-point of the resummed distribution to match the end-point of the fixed-order computation.
In order to accomplish this we will make use of the techniques described in~\cite{Catani:1992ua,Jones:2003yv}. The main point is to modify the
argument of the logarithms of $\tau$ to:
\begin{equation}
\log\left(x_{L}\tau\right)\to-\log\left(\frac{1}{x_{L}\tau}-\frac{1}{x_{L}\tau_{\mathrm{max}}}+1\right)=\log\left(\frac{x_{L}\tau\tau_{\mathrm{max}}}{\tau_{\mathrm{max}}-\tau+x_{L}\tau\tau_{\mathrm{max}}}\right)=\log\bar{\tau}.
\end{equation}
This modification is enough to to reduce the resummation to $0$ at the end-point $\tau_{\mathrm{max}}$. However, the expansion has an additional constant:
\begin{equation}\label{eq:expansion-plain}
\tau\frac{d\sigma_{\mathrm{\mathrm{exp}}}}{d\tau}=\frac{\alpha_{\mathrm{s}}}{\pi}\left[\frac{1}{2}G_{12}\log\bar{\tau}+G_{11}\right],
\end{equation}
where $G_{ij}$ indicates the different coefficients of the expansion. With the new modified logarithms that go to $0$ at the end-point this expression becomes equal to $G_{11}$ at the end-point. In order to also make the value of the expansion at the end-point equal to $0$ we will add a term $-G_{11}\tau/\tau_{\mathrm{max}}$ to this expression resulting
in:
\begin{equation}
\tau\frac{d\sigma_{\mathrm{\mathrm{exp}}}}{d\tau}=\frac{\alpha_{\mathrm{s}}}{\pi}\left[\frac{1}{2}G_{12}\log\bar{\tau}+G_{11}\left(1-\frac{\tau}{\tau_{\mathrm{max}}}\right)\right].
\end{equation}
This does mean that the resummation need to include an additional term
\begin{equation}
\Sigma\(\tau\)\to \Sigma\(\tau\)\exp\left[-\frac{\as}{\pi}\frac{\tau}{\tau_{\mathrm{max}}}G_{11}\log\bar{\tau}\right],
\end{equation}
in order to ensure that the expansion of exponential reproduces the correct result. For the derivative of $R$ we will suppress any power corrections that are unnecessary to make both resummation and expansion approach $0$ at the end-point:
\begin{equation}
\tau\frac{d\sigma_{\mathrm{res}}}{d\tau}=\left(F^{\prime}\left(\log\bar{\tau}\right)-\frac{\as}{\pi}\frac{\tau}{\tau_{\mathrm{max}}}G_{11}\right)C\exp\left[F\left(\log\bar{\tau}\right)-\frac{\as}{\pi}\frac{\tau}{\tau_{\mathrm{max}}}G_{11}\log\bar{\tau}\right].
\end{equation}
where we have taken the derivative with respect to $\log\bar{\tau}$ instead of $\log\tau$. All of these modifications are power suppressed terms and do not alter the $\tau\to0$ limit.

Strictly speaking, the end-point modification is relevant only for the ungroomed part of the spectrum. However, groomed and ungroomed distributions should line up at the transition point and it is therefore convenient to also modify the soft-drop distribution.
Because logarithms of $z_{\mathrm{cut}}$ become logarithms of $\tau$ beyond the transition point, we treat them on a equal footing as the logarithms of $\tau$.
Therefore, we introduce a variant of our resummed expression, which only differs by power corrections
\begin{eqnarray}
\tau\frac{d\sigma_{\mathrm{res}}}{d\tau}&=&\left(F^{\prime}\left(\log\bar{\tau},\log z_{\mathrm{cut}}\right)-\frac{\as}{\pi}\frac{\tau}{\tau_{\mathrm{max}}}\left(G^{\mathrm{(SD)}}_{11}+S_{0}^{\prime}\right)\right)C^{\mathrm{(SD)}}\(\frac{2\tau}{\zc}\)\exp\left[F\left(\log\bar{\tau},\log z_{\mathrm{cut}}\right)\right]\nonumber\\
&&\times\prod_{K}\exp\[\frac{\as}{\pi}\frac{\tau}{\tau_{\mathrm{max}}}\frac{G_{11}^{(K)}}{p^{\(\bar{N}\)}_K}\(-p^{\(\bar{N}\)}_K\log\bar{\tau}+p^{\(\zc\)}_K\log\(x_L\frac{\zc}{2}\)\)\],
\end{eqnarray}
where $S_0'$ is the derivative of the dilogarithmic contribution, $p^{(i)}_{K}$ are the powers of a variable $i$ in the scale of a function $K$ and $C^{\mathrm{(SD)}}$ depends on $2\tau/\zc$ through means of the dilogarithm and reduces to $C$ at the transition point. The coefficient $G^{\mathrm{(SD)}}_{11}$ originates from the expansion as described in the case of ungroomed thrust, Eq.~(\ref{eq:expansion-plain}), which is equal to the sum over all possible functions $K$ for $G_{11}^{(K)}$. 
Here the power-suppressed term for the dilogarithm itself $C^{\mathrm{(SD)}}$ was not included in order to insure that $C^{\mathrm{(SD)}}(1)=C$. These power corrections become equal to the end point corrections for ungroomed thrust at the transition point.

\phantomsection
\addcontentsline{toc}{section}{References}
\bibliographystyle{jhep}
\bibliography{references}
\end{document}